\documentclass[%
 reprint,
 superscriptaddress,
nofootinbib,
 amsmath,amssymb,
 aps,
]{revtex4-1}

\usepackage{graphicx}% Include figure files
\usepackage{dcolumn}% Align table columns on decimal point
\usepackage{bm}% bold math
\usepackage{hhline}
\usepackage{xcolor}
\usepackage{siunitx}
\usepackage{hyperref}% add hypertext capabilities
\usepackage{ulem}
\usepackage{orcidlink}
\newcommand{\chieff}{\chi_{\mathrm{eff}}}

\ExplSyntaxOn
\NewDocumentCommand{\longdash}{ O{2} }
 {
  --\prg_replicate:nn { #1 - 1 } { \negthinspace -- }
 }
\ExplSyntaxOff

\newcounter{RunIDCounter}

\newcommand{\ridref}[1]{\refstepcounter{RunIDCounter}\label{#1}\theRunIDCounter}

\definecolor{shgreen}{rgb}{0.15625, 0.609375, 0.316406}

\begin{document}

%\preprint{APS/123-QED}

\title{The anti-aligned spin of GW191109: glitch mitigation and its implications}

\author{Rhiannon Udall~\orcidlink{0000-0001-6877-3278}}
\email{rudall@caltech.edu}
\affiliation{%
Department of Physics, California Institute of Technology, Pasadena, California 91125, USA
}%
\affiliation{
LIGO Laboratory, California Institute of Technology, Pasadena, California 91125, USA
}
\author{Sophie Hourihane~\orcidlink{0000-0002-9152-0719}}
\email{sohour@caltech.edu}
\affiliation{%
Department of Physics, California Institute of Technology, Pasadena, California 91125, USA
}
\affiliation{
LIGO Laboratory, California Institute of Technology, Pasadena, California 91125, USA
}
\author{Simona Miller~\orcidlink{0000-0001-5670-7046}}
\email{smiller@caltech.edu}
\affiliation{%
Department of Physics, California Institute of Technology, Pasadena, California 91125, USA
}
\affiliation{
LIGO Laboratory, California Institute of Technology, Pasadena, California 91125, USA
}

\author{Derek Davis~\orcidlink{0000-0001-5620-6751}}
\email{dedavis@caltech.edu}
\affiliation{%
Department of Physics, California Institute of Technology, Pasadena, California 91125, USA
}
\affiliation{
LIGO Laboratory, California Institute of Technology, Pasadena, California 91125, USA
}
\author{Katerina Chatziioannou~\orcidlink{0000-0002-5833-413X}}
\email{kchatziioannou@caltech.edu}
\affiliation{%
Department of Physics, California Institute of Technology, Pasadena, California 91125, USA
}
\affiliation{
LIGO Laboratory, California Institute of Technology, Pasadena, California 91125, USA
}
\author{Max Isi~\orcidlink{0000-0001-8830-8672}}
\email{misi@flatironinstitute.org}
\affiliation{%
Center for Computational Astrophysics, Flatiron Institute, New York, NY 10010, USA
}
\author{Howard Deshong~\orcidlink{0000-0002-3473-9239}}
\email{hdeshong@caltech.edu}
\affiliation{
Schmidt Academy for Software Engineering, Pasadena, California 91125, USA
}
\affiliation{
LIGO Laboratory, California Institute of Technology, Pasadena, California 91125, USA
}

\date{\today}% It is always \today, today,

\begin{abstract}

    With a high total mass and an inferred effective spin anti-aligned with the orbital axis at the 99.9\% level, GW191109 is one of the most promising candidates for a dynamical formation origin
    among gravitational wave events observed so far.
    However, the data containing GW191109 are afflicted with terrestrial noise transients, i.e., detector glitches, generated by the scattering of laser light in both LIGO detectors.
    We study the implications of the glitch(es) on the inferred properties and astrophysical interpretation of GW191109.
    Using time- and frequency-domain analysis methods, we isolate the critical data for spin inference to
    $35-40 \, \si{Hz}$ and $0.1 - 0.04 \, \si{s}$ before the merger in LIGO Livingston, directly coincident with the glitch.
    Using two models of glitch behavior, one tailored to slow scattered light and one more generic, we perform joint inference of the glitch and binary parameters.
    When the glitch is modeled as slow scattered light, the binary parameters favor anti-aligned spins, in agreement with existing interpretations.
    When more flexible glitch modeling based on sine-Gaussian wavelets is used instead, a bimodal aligned/anti-aligned solution emerges. The anti-aligned spin mode is correlated with a weaker inferred glitch and preferred by $\sim70:30$ compared to the aligned spin mode and a stronger inferred glitch.
    We conclude that if we assume that the data are only impacted by slow scattering noise, then the anti-aligned spin inference is robust.
    However, the data alone cannot validate this assumption and resolve the anti-aligned spin and potentially dynamical formation history of GW191109.
\end{abstract}

\maketitle

%%%%%%%%%%%%%%%%%%%%%%%%%%%%%%%%%%
\section{Introduction}
%%%%%%%%%%%%%%%%%%%%%%%%%%%%%%%%%%

Reported in the third gravitational wave (GW) transient catalog (GWTC-3)~\cite{KAGRA:2021vkt}, GW191109\_010717 (more concisely GW191109) stands out among existing binary black hole (BBH) signals. 
With source-frame primary and secondary masses of $m_1=65^{+11}_{-11}\,M_\odot$ and $m_2=47^{+15}_{-13}\,M_\odot$ (90\% symmetric credible intervals), it is among the most massive events.
Furthermore, there is significant support for black hole (BH) spins anti-aligned with the orbital angular momentum: the mass-weighted effective spin~\cite{Racine:2008qv,Ajith:2009bn, Santamaria:2010yb} is $\chieff = -0.29^{+0.42}_{-0.31}$.
For these reasons, as well as support for unequal masses, $q=m_2/m_1=0.73^{+0.21}_{-0.24}$, spin-precession, and hints of eccentricity~\cite{Romero-Shaw:2022xko, Gupte:2024jfe}, the binary is potentially of dynamical and/or hierarchical origin~\cite{Antonelli:2023gpu, Zhang:2023fpp} and impacts population inference~\cite{Tong:2022iws, Adamcewicz:2023mov}. 

Multiple GW191109 properties hint toward a dynamical origin.
High masses, above the pair-instability supernova (PISN) limit of $45-70\,M_\odot$ (depending on modeling assumptions)~\cite{Farmer:2019jed, Woosley:2021xba}, may require a hierarchical mechanism in order to form and merge.
Asymmetric masses, in particular, might imply the merger of a second- and a first-generation BH~\cite{Antonelli:2023gpu}.
Furthermore, population synthesis simulations of isolated formation scenarios typically find little support for spins anti-aligned with the orbital angular momentum, unless supernova kicks are exceptionally high~\cite{Kalogera:1999tq, Zhang:2023fpp,  Gerosa:2018wbw}.
Finally, eccentricity would also be challenging to explain except by dynamical processes~\cite{Zevin:2021rtf, Romero-Shaw:2022xko, Romero-Shaw:2021ual}, due to the rapid orbit circularization by GW emission~\cite{Peters:1964zz}.
 
Given their astrophysical implications, the inferred properties of GW191109 are worth scrutinizing.
The first potential source of systematics is the waveform used to model the signal. 
GWTC-3 employed the \textsc{IMRPhenomXPHM}~\cite{Pratten:2020ceb} and \textsc{SEOBNRv4PHM} approximants~\cite{Ossokine:2020kjp}, with inference performed by \textsc{Bilby}~\cite{bilby_paper, bilby_pipe_paper} and \textsc{RIFT}~\cite{Lange:2018pyp} respectively.
Both models include the physical effects of higher-order modes and spin-precession, and headline results (as quoted above) are their average. 
However, GW191109 is flagged for systematic differences between approximants~\cite{KAGRA:2021vkt}, especially for the binary inclination (edge-on versus face-on/off respectively) and the longer $\chieff>0$ tail with \textsc{IMRPhenomXPHM}. 
A third waveform, \textsc{NRSur7dq4}~\cite{Varma:2019csw}, was employed in Ref.~\cite{Islam:2023zzj}.
A direct surrogate of numerical relativity simulations, \textsc{NRSur7dq4} is expected to be the most accurate available model for systems with high masses and spins~\cite{Varma:2019csw,Hannam:2021pit, Islam:2023zzj}.
These results bolster the evidence for dynamical origin, with a more negative spin, $\chieff=-0.38^{+0.21}_{-0.20}$, asymmetric masses, $q=0.65^{+0.20}_{-0.19}$, and a precessing spin parameter~\cite{Schmidt:2014iyl} of $\chi_p=0.59^{+0.26}_{-0.27}$. 
While waveform systematics remain relevant, the broad agreement between three waveforms (including a direct surrogate to numerical relativity) that $\chieff\lesssim0$ to varying credibility, suggests that subsequent interpretations of its formation history remain valid. 

A second potential source of systematics concerns modeling the detector noise.
Around GW191109's arrival, both LIGO~\cite{LIGOScientific:2014pky} detectors experienced a terrestrial noise transient known as a scattered light glitch~\cite{KAGRA:2021vkt, Davis:2022ird, Udall:2022vkv}.
The Virgo detector~\cite{Acernese:2015gua} was offline at this time, and so only the LIGO detectors contributed to the observation.
In LIGO Hanford (LHO), the glitch power was at a nadir while the event was in the detection band, making its impact on the inferred parameters negligible, see App.~\ref{app:lho-glitch}.
As such, we ignore the LHO glitch going forward.
By contrast, glitch power in the Livingston detector (LLO) was directly coincident in time and frequency with the signal, a circumstance which could bias astrophysical inference~\cite{Pankow:2018qpo,Chatziioannou:2021ezd,Payne:2022spz, Hourihane:2022doe, Ghonge:2023ksb}.
Specifically, glitch power extends up to $\sim40\,$Hz, coincident with the signal, see Fig.~\ref{fig:witness-spectrogram-overlay}.
Spin parameters might be particularly susceptible to such data quality issues due to the relatively smaller imprint they leave on signals compared to, e.g., the BH masses.
For example, GW200129 shows evidence of spin-precession~\cite{KAGRA:2021vkt,Hannam:2021pit}, but its significance depends on how the glitch that overlapped that signal is modeled~\cite{Payne:2022spz, Macas:2023wiw}.
%Since GW191109 has the most negative inferered posterior of $\chieff$ in GWTC-3, and this spin measurement is the primary quality which distinguishes GW191109 from other heavy systems with relatively uninteresting spins, confident astrophysical understanding of this event necessarily requires robust understanding of the noise in the detector. 

The headline GWTC-3 results were obtained after an estimate for the glitch had been subtracted from the data. 
The two-step process involved first modeling the signal and the glitch with a flexible sum of coherent and incoherent wavelets respectively with \textsc{BayesWave}~\cite{Cornish:2014kda,Littenberg:2014oda, Cornish:2020dwh}.
Second, a fair draw from the glitch posterior was subtracted and the system parameters were inferred as quoted above.
This procedure has been shown to generally lead to unbiased mass and (aligned) spin inference~\cite{Pankow:2018qpo, Hourihane:2022doe}. However, uncertainties remain related to \textsc{BayesWave}'s glitch model and in the fair draw chosen to be subtracted.
These effects were investigated in Ref.~\cite{Davis:2022ird}, albeit with a simpler waveform model with single-spin precession and no higher-order modes,  \textsc{IMRPhenomPv2}~\cite{Hannam:2013oca}.
Glitch mitigation was found to affect the $\chieff$ inference by a similar amount as waveform systematics.
%($\chieff=-0.43^{+0.28}_{-0.27}$ before removal of the glitch realization to $\chieff=-0.34^{+0.20}_{-0.26}$ after removal).
Completely removing the glitch-affected data, i.e.~all LLO data below 40\,Hz, instead resulted in a dramatic shift of $\chieff$ to positive values $\chieff=0.27^{+0.24}_{-0.48}$.

The stark impact of glitch-affected data on astrophysically-impactful spin inference motivates our study.
In Sec.~\ref{sec:why-is-the-inferred-chieff-negative} we extend Ref.~\cite{Davis:2022ird} to explore the manner in which the data inform the system parameters.
Using \textsc{NRSur7dq4} and a frequency-domain analysis, we find that the LLO data between $30$ and $40\,\text{Hz}$ are crucial for spin inference: excluding $30-40$\,Hz data shifts the probability of $\chieff<0$ from 99.4\% to 32.2\%, effectively wiping out any preference for for anti-aligned spins.
%, with shifts from $\chieff=-0.25^{+0.20}_{-0.21}$ when analyzing with data corresponding to frequencies $f\geq 20 \text{Hz}$ and $\chieff=-0.22^{+0.21}_{-0.22}$ when analyzing with data corresponding to frequencies $f\geq 30 \text{Hz}$ to $\chieff=0.18^{+0.24}_{-0.29}$ when analyzing with data corresponding to frequencies $f\geq 40 \text{Hz}$.
A similar time-domain analysis~\cite{Miller:2023ncs} highlights the role of the data $0.1-0.04\,$s prior to merger.
%We also perform an analogous test in the time domain, and find similar shifts, with \RU{Update these with final TD result when available}$\chieff=-0.24^{+0.33}_{-0.29}$ when the full data are used, $\chieff=-0.19^{+0.30}_{-0.30}$ when only data from $0.1 \text{s}$ before merger onwards are used, and $\chieff=0.1^{+0.29}_{-0.4}$ when only data from $0.04 \text{s}$ before merger onwards are used.
These data, which inform the $\chieff<0$ measurement, coincide in time and frequency with excess power in LLO, see Fig.~\ref{fig:TimeFrequencyCombined} and in particular the excess power at $\sim36\,$Hz.
To check whether such dramatic shifts in support for $\chieff<0$ are possible from Gaussian noise alone, we analyze $100$ simulated signals consistent with GW191109.
We find that shifts of this magnitude are unlikely but not impossible as $6\%$ of the simulations experience a larger shift than GW191109. 

In Sec.~\ref{sec:application-of-glitch-mitigation}, we focus on the $36\,$Hz excess power and address the key question: is the excess power part of the signal (and hence $\chieff<0$) or is it part of the glitch (and hence inference has been affected by systematics)?
Rather than the two-step process of glitch fitting and subtraction, we perform a full analysis where we \textit{simultaneously} model both the signal and the glitch.
Using a physically motivated model for scattered light glitches~\cite{Udall:2022vkv} we find $\chieff<0$ at the $99.9\%$ level using \textsc{NRSur7dq4}. 
We attribute this to the fact that the $36\,$Hz power is more contained in time than expected for scattered light glitches that are characterized by extended arches in time-frequency.
This analysis, therefore, attributes the $36\,$Hz power to the signal and thus prefers $\chieff<0$.
It is, however, possible that not all terrestrial power is due to scattered light or that the physical model of Ref.~\cite{Udall:2022vkv} does not capture all scattered light power.
Instead, using a more flexible model for the glitch based on wavelets and \textsc{BayesWave} and \textsc{IMRPhenomXPHM} we obtain a bimodal solution for the spin. 
One mode, preferred at the $70:30$ level, attributes most of the $36\,$Hz power to the signal and results in $\chieff<0$. 
The second mode attributes this power to the glitch and results in $\chieff>0$.
Given the low signal-to-noise ratio (SNR) of the $36\,$Hz power, these results are impacted by the priors of the glitch model parameters at the few percent level.

In Sec.~\ref{sec:conclusion} we summarize our conclusions.
Physically grounded assumptions about the behavior of scattered light glitches lend support to $\chieff<0$ for GW191109, and thus a dynamical origin.
However, both systematic limitations on scattered light models and statistical uncertainty due to low SNR of the excess power and the impact of glitch priors prevent us from making that determination confidently.
While the crucial $36\,$Hz power is not part of the scattered light glitch as modeled in Ref.~\cite{Udall:2022vkv}, we cannot rule out glitch mismodeling or other types of terrestrial noise.

%%%%%%%%%%%%%%%%%%%%%%%%%%%%%%%%%%
\section{Modeling signals and glitches}\label{sec:methodology}
%%%%%%%%%%%%%%%%%%%%%%%%%%%%%%%%%

The relevant data contain the GW191109 signal, glitch power, and Gaussian noise. In this section, we describe how we model the signal (Sec.~\ref{sec:modeling-the-compact-binary}), the glitch (Sec.~\ref{sec:glitch-modeling}), and methods for glitch mitigation (Sec.~\ref{sec:glitch-mitigation-approaches}).
We focus on the respective strengths and weaknesses of each approach and what unique information each supplies.
All analyses model the Gaussian noise component with the power spectral densities (PSDs) from the GWTC-3 data release~\cite{KAGRA:2023pio, ligo_scientific_collaboration_and_virgo_2023_8177023}.
% Citing GWOSC paper rather than glitch modeling since PSDs technically came from
% PE Results pages
%~\cite{ligo_scientific_collaboration_and_virgo_2021_5546680}.
Detailed settings and identification numbers for all analyses are given in Table~\ref{tab:all-analyses-with-settings} in App.~\ref{app:settings}.

%%%%%%%%%%%%%%%%%%%%%%%%%%%%%%%%%%%%%
\subsection{Modeling the Compact Binary Signal}\label{sec:modeling-the-compact-binary}

We use both time- and frequency-domain techniques to model the signal with either waveform approximants for compact binary signals or, more generically, with sine-Gaussian wavelets.
All analyses consider data surrounding the nominal trigger time of GW191109,  GPS time $1257296855.22$, and employ a sampling rate of $1024 \, \si{Hz}$, with the maximum analysis frequency set to $7/8$ of the Nyquist frequency.
Unless otherwise noted, analyses that model only the compact binary (and not the glitch) use a minimum frequency of $20 \, \si{Hz}$ in both detectors. 
We use standard compact-binary priors~\cite{bilby_pipe_paper}, notably uniform in detector-frame component masses and spin magnitude and orientation.

%%%%%%%%%%%%%%%%%%%%%%%%%%%%%%%%%%%%%
\subsubsection{Frequency-domain inference}\label{sec:fd-inference-methodology}
%%%%%%%%%%%%%%%%%%%%%%%%%%%%%%%%%%%%%

Frequency domain analyses with waveform approximants are based on \textsc{Bilby}~\cite{bilby_paper,bilby_pipe_paper} with its implementation of the \textsc{dynesty} sampler~\cite{Speagle:2019ivv} and \textsc{BayesWave}~\cite{Chatziioannou:2021ezd}, both analyzing 4\,s of data.
The former models the signal with \textsc{NRSur7dq4}~\cite{Varma:2019csw} and the latter with \textsc{IMRPhenomXPHM}~\cite{Pratten:2020ceb} (though for consistency we also perform checks with the former using \textsc{IMRPhenomXPHM} in App.~\ref{sec:bilby-imrx}).
\textsc{NRSur7dq4} supports a minimum mass ratio of $0.25$ and minimum detector-frame chirp mass of $35\, M_\odot$; neither restriction affects the analysis. 
We extend into the extrapolation region in spins, setting a maximum spin magnitude of $0.99$. 
For comparison, we also perform analyses with \textsc{BayesWave} where the signal is modeled as a flexible sum of coherent sine-Gaussian wavelets~\cite{Cornish:2014kda,Cornish:2020dwh}. 
Settings are similar to the glitch wavelet analysis described in Sec.~\ref{sec:glitch-modeling-bayeswave}, only here, the wavelets are coherently projected across the two detectors rather than being independent.

%%%%%%%%%%%%%%%%%%%%%%%%%%%%%%%%%%%%%
\subsubsection{Time-domain inference}\label{sec:td-inference-methodology}
%%%%%%%%%%%%%%%%%%%%%%%%%%%%%%%%%%%%%

While GW inference is typically conducted in the frequency domain for computational efficiency, it can equivalently be conducted in the time domain~\cite{Isi:2021iql,Isi:2020tac,Isi:2019aib,Carullo:2019flw}. 
Frequency domain analyses are non-local in time; to avoid non-trivial likelihood modifications~\cite{Capano:2021etf}, time-domain inference is necessary in order to isolate purely temporal features of the data. 
Below, we truncate the GW191109 data at different times around the 36\,Hz excess power, and independently conduct inference on the pre- or post-cutoff-time data. 
For this, we use the time-domain inference code employed in Ref.~\cite{Miller:2023ncs} to study the GW190521 properties and which was based on time-domain implementations targeting post-merger data~\cite{Isi:2019aib,Isi:2020tac}. 
All time-domain results are based on regions of 1\,s of data around GW191109's trigger time and employ \textsc{NRSur7dq4}~\cite{Varma:2019csw}.
The same PSDs are used in the time domain analyses are the same as those in the frequency domain analyses, i.e.~ from the GWTC-3 data release~\cite{KAGRA:2023pio}.

%%%%%%%%%%%%%%%%%%%%%%%%%%%%%%%%%%%%%
\subsection{Modeling the Glitch}\label{sec:glitch-modeling}
%%%%%%%%%%%%%%%%%%%%%%%%%%%%%%%%%%%%%

Both LHO and LLO experienced slow scattering noise around the time of GW191109.
We use two models for the glitch power: a physically motivated model tailored to slow scattering, implemented in \textsc{Bilby}, and a more flexible wavelet model, implemented in \textsc{BayesWave}.

%%%%%%%%%%%%%%%%%%%%%%%%%%%%%%%%%%%%%
\subsubsection{Physically-parameterized scattering}\label{sec:scattered-light-modeling}
%%%%%%%%%%%%%%%%%%%%%%%%%%%%%%%%%%%%%

\begin{figure*}
    \centering
    \includegraphics[width=\textwidth]{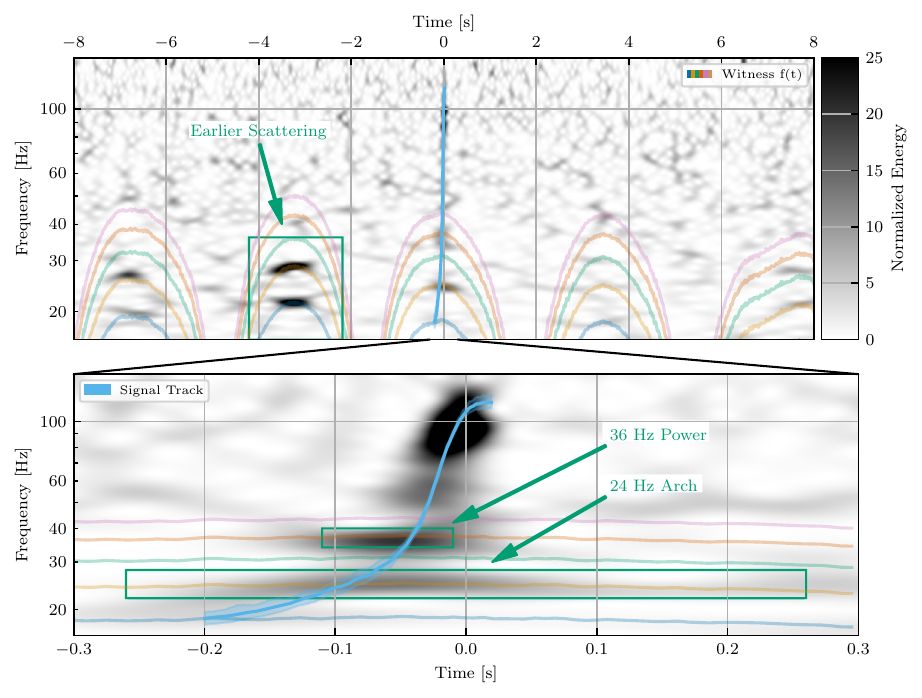}
    \caption{
    Spectrograms of the original (before glitch mitigation) data in LLO centered around the time of GW191109.
    The top panel shows $\pm 8 \, \si{s}$ of data, while the bottom panel zooms in around the event.
    Onto this, we plot the time-frequency tracks of the scattered light glitch, as predicted by the motion observed in the witness channel \texttt{L1:SUS-ETMX\_L2\_WIT\_L\_DQ}.
    This is the witness to the penultimate stage of the reaction chain pendulum for the X-arm end test mass.
    The scattering surface is the final stage of the reaction chain, and so this witness does not perfectly capture the motion of the scattering surface; to compensate, we apply a static coefficient of $1.38$ to the predicted frequency, such that it is calibrated to the prominent scattering arches $\sim3 \, \si{s}$ before the event.
    We also plot the inferred signal from a \textsc{NRSur7dq4} analysis of full-bandwidth data after glitch subtraction (Run~\ref{rid:subtracted-fd-fL20-fH20} in Table~\ref{tab:all-analyses-with-settings}).
    We annotate three regions of interest: the prominent scattering before the event (top panel), the long-duration excess power at $24 \, \si{Hz}$ (bottom panel), and the short-duration excess power at $36 \, \si{Hz}$ (bottom panel).
    Both the 24\,Hz and the 36\,Hz excess power coincide with expected glitch arches, however only the former has an arch-like shape.
    }
    \label{fig:witness-spectrogram-overlay}
\end{figure*}

%An alternate approach to glitch modeling, which was introduced for use in parameter estimation by \cite{Udall:2022vkv}, may be applied for scattered light glitches, which are known to be present in both detectors during GW191109. 
As the name implies, scattered light glitches arise due to laser light that scatters off the main beam path, bounces off a surface, and recombines with the main beam~\cite{Accadia:2010zzb, LIGO:2020zwl, LIGO:2021ppb, Udall:2022vkv, Tolley:2023umc}.
During periods of significant ground motion when the scattering surface moves, this light acquires a phase offset, resulting in excess noise. 
Figure~\ref{fig:witness-spectrogram-overlay} shows a spectrogram of the LLO data, along with the frequency tracks of the scattering excess noise as predicted by a witness data stream that captured the motion of the suspected scattering surface.
The effect of scattering is most easily discernible 3\,s before the signal, taking the form of a ``stack" of arches, which is characteristic of slow scattering. 
Slow scattering results from low-frequency ground motion, ${\sim} 0.05 {-} 0.3 \, \si{Hz} $, driving slow movement of the scattering surface~\cite{LIGO:2020zwl}.
This induces phase noise with frequency~\cite{Accadia:2010zzb}
\begin{equation}
    f(t) = \biggr{|} \frac{2 v_{sc}(t)}{\lambda}\biggr{|}\,,
\end{equation}
with $v_{sc}(t)$ being the velocity of the scattering surface and $\lambda=1064 \, \si{nm}$ is the wavelength of the laser.
In order for the glitch frequency to reach the analysis band, the scattered light must bounce multiple times, yielding a fixed frequency ratio between arches as the same amount of phase offset is accumulated with each successive bounce.

This picture forms the basis for a parametrized model for slow scattering that treats the scattering surface as a simple harmonic oscillator.
We use the physically parameterized scattering model proposed in Ref.~\cite{Udall:2022vkv}.
The model is a sum of frequency-modulated sinusoids with $2N + 4$  parameters, where $N$ is the number of arches:
\begin{equation}
        g(t) = \sum_{k=0}^N A_k \sin\biggr{[}
        \frac{f_{h, 0} + k \delta f_{h}}{f_{\rm mod}} \sin\biggr{(}2 \pi f_{\rm mod} (t - t_{c})\biggr{)} + \phi_k \biggr{]}\,.
\end{equation}
The peak frequency of the lowest arch is $f_{h,0}$ and the spacing in peak frequencies between adjacent arches is $\delta f_{h}$, such that the peak frequency of the $k$th arch is $f_{h, 0} + k \delta f_{h, 0}$.\footnote{Unlike Ref.~\cite{Udall:2022vkv}, we fix the frequency ratio between arches to $\delta f_{h, 0}$, thus eliminating $N-1$ parameters.} 
The modulation frequency $f_{\rm mod}$ corresponds to the motion of the scattering surface (and hence the driving ground motion) and sets the width of the arch, while $t_c$ is the time of peak frequency.
Each arch $k$ further has an independent amplitude $A_k$ and phase $\phi_k$.

Priors on these parameters reflect the physical slow scattering picture.
For $\delta f_{h}$ and $f_{h, 0}$, we place uniform priors around the approximate values read from Fig.~\ref{fig:witness-spectrogram-overlay}, $\delta f_{h} \sim \mathcal{U}(5, 8)\,$Hz and $f_{h, 0} \sim \mathcal{U}(18, 20)\,$Hz, while for $\phi_k$ we set a uniform periodic prior, $\phi_k \sim \mathcal{U}(0, 2 \pi)$.
We employ two sets of priors on $f_{\rm mod}$.
``Physical" priors limit the modulation frequency to the microseism band $f_{\rm mod}\sim \mathcal{U}(0.05 - 0.3) \, \si{Hz} $~\cite{LIGO:2020zwl}. 
``Targeted" priors further restrict the modulation based on the witness motion $f_{\rm mod}\sim \mathcal{U}(0.05 - 0.15) \, \si{Hz}$.
While the former choice is more agnostic, the latter maximizes information from witness channels.
%\RU{Maybe we'll explore this in the appendix, otherwise take these sentences out} One may also set ``unphysical" priors which allow ground motion frequencies greatly exceeding the observed motion.
%This form of the model is much more flexible (high modulation frequency arches come to resemble wavelets), but loses the physical motivation which grounds this model; nonetheless, we'll explore these briefly in Appendix \ref{app:other-glitch-models}.
Since the detector sensitivity varies by orders of magnitude in the frequency region spanned by the arches, we explore both a uniform and log-uniform amplitude for the amplitude $A_k$.
We do not impose a relation between the arch amplitudes; while amplitudes might be expected to decrease with each arch, this is not universally the case~\cite{Udall:2022vkv}.
%Next, we may choose to adopt either uniform or log-uniform priors on the amplitudes of respective arches.
%Because the sensitivity of the detector varies by orders of magnitude over the frequency region in which the arches sit, arches may contribute similar SNR but have orders of magnitude variation in their amplitudes.
%We adopt the same minimum and maximum amplitude in the prior for each arch, due to the potential subtleties of inferring any more relationship (while amplitudes are expected to decrease with each arch, this is not universally the case \cite{Udall:2022vkv}).
%Log-uniform priors allow better differentiation over orders of magnitude, but induce a strong prior preference for weaker arches.

The number of arches $N$ is fixed and not a parameter of the model that is varied, unlike the flexible glitch model with \textsc{BayesWave} discussed in Sec.~\ref{sec:glitch-modeling-bayeswave}.
%Unlike \textsc{BayesWave}, this must be fixed as part of the configuration of the run, and so (especially in uniform amplitude priors) it is possible to spuriously introduce more arches than should be present. 
The choice of the number of arches, therefore, impacts the results, especially for the uniform amplitude prior.
Motivated by Fig.~\ref{fig:witness-spectrogram-overlay} we set $N=5$, a choice which we investigate in App.~\ref{sec:appendix-effect-of-glitch-priors}. 
All analyses that model the glitch with the slow scattering model further employ a reduced minimum frequency of $16 \, \si{Hz}$ in LLO.
Though the signal SNR, $\rho$, is negligible between $16$ and $20 \, \si{Hz}$ (0.16\% of $\rho^2$ in LLO), this setting accommodates the $\sim 18 \, \si{Hz}$ arch, which in turn informs the upper arches. 

If the glitch overlapping GW191109 is consistent with the physical picture that motivates the slow scattering model, corresponding analyses provide the most sensitive results on the system properties.
However, the model is also restricted to an interpretation of slow scattering and does not provide a means to test this assumption.
If other non-Gaussian transient noise is present or if the physical picture does not fully capture the glitch morphology, biases might arise.

%%%%%%%%%%%%%%%%%%%%%%%%%%%%%%%%%%%%%
\subsubsection{Wavelet glitch model}\label{sec:glitch-modeling-bayeswave}
%%%%%%%%%%%%%%%%%%%%%%%%%%%%%%%%%%%%%

To mitigate against glitch modeling systematics, we also employ a more flexible approach with \textsc{BayesWave} which models transient, non-Gaussian noise independently in each detector as sums of sine-Gaussian, Morlet-Gabor wavelets~\cite{Cornish:2014kda,Cornish:2020dwh}. 
Such wavelets are an overcomplete basis and any smooth function can be described with some linear combination of wavelets. 
Thus, this glitch model is flexible enough to fit a wide range of non-Gaussian transients without fine-tuning, including slow scattering~\cite{Chatziioannou:2021ezd,Hourihane:2022doe}.
Unlike the parameterized scattering model, the \textsc{BayesWave} glitch model is purely phenomenological, though motivated by the generic morphology of the LIGO glitches. 
Each wavelet is described by 5 parameters: central time $t$ and frequency $f$, quality factor $Q$ describing how quickly it is damped, amplitude $A$, and phase $\phi$. 
We employ uniform priors over all parameters other than the amplitude, which is set through a prior on the wavelet SNR that peaks at $5$~\cite{Cornish:2014kda}.
In addition to these parameters, the number of wavelets in each detector is also a variable and sampled over with a uniform prior. 
Uniform prior bounds are wide enough so as to not affect the posterior.

%In this study, two different models are used to fit the signal in the data. The first model, called the gw-wavelet model, represents un-modeled coherent power between the detectors. It is similar to the glitch model in that it is a sum of wavelets, but with additional constraints: 1) the same wavelets are present in all detectors, and 2) the wavelets must be coherent across the detector network as a genuine astrophysical signal would be. 

% \begin{table}[h!]
% \centering
% \begin{tabular}{|>{\raggedright\arraybackslash}m{4cm}|>{\raggedright\arraybackslash}m{4cm}|}
% \hline
% \textbf{Parameters} & \textbf{Prior} \\ \hline
% time & U(0, 1) \\ \hline
% frequency & U(x, x) \\ \hline
% Q & U(x, x) \\ \hline
% \end{tabular}
% \caption{Parameters and their Priors}
% \label{table:params_priors}
% \end{table}

%%%%%%%%%%%%%%%%%%%%%%%%%%%%%%%%%
\subsection{Glitch Mitigation Approaches}\label{sec:glitch-mitigation-approaches}
%%%%%%%%%%%%%%%%%%%%%%%%%%%%%%%%%%%%%

We employ three approaches to mitigate and study the impact of the glitch on inference: (1) discarding the affected data, (2) subtracting an estimate for the glitch from the data, and (3) simultaneously modeling the signal and glitch and obtaining source parameters for the former by marginalizing over the latter.

%%%%%%%%%%%%%%%%%%%%%%%%%%%%%%%%%%%%%
\subsubsection{Discarding Affected Data}\label{sec:removal-of-affected-data}
%%%%%%%%%%%%%%%%%%%%%%%%%%%%%%%%%%%%%

The most straightforward way to mitigate the impact of a glitch is to discard the affected data, either by band-passing in the frequency domain or by analyzing limited segments in the time domain~\cite{Pankow:2018qpo,Davis:2022ird, Payne:2022spz}. 
%Since most glitches occur in the lower frequency regions of the detector, and all analyses must already apply some band-passing, this has been applied in the past in a number of cases .
%While explicit time domain inference of the type has not been applied to this problem before, a related gating procedure is employed by search pipelines, including the immediate followup to GW170817 \cite{Pankow:2018qpo}.
While straightforward to implement,
%These methods are conceptually simple and allow us to be sure that the effects of glitch are fully removed from the analysis.
such methods forego all information in the discarded data, making them suboptimal. 
We instead follow Refs.~\cite{Davis:2022ird, Payne:2022spz} and discard glitch-affected data only as a consistency check and to study the impact of the glitch, or its residual, on inference.
Such analyses confirm that mitigation is necessary and provide insights into the detailed behavior of the data.

%%%%%%%%%%%%%%%%%%%%%%%%%%%%%%%%%%%%%
\subsubsection{Subtraction of a Glitch Estimate}
%%%%%%%%%%%%%%%%%%%% %%%%%%%%%%%%%%%%%

GWTC-3 results on GW191109 were obtained after an estimate of the glitch was subtracted from the data~\cite{KAGRA:2021vkt}.
In most cases, the estimate for the glitch is a fair draw from a previous analysis with \textsc{BayesWave}~\cite{LIGOScientific:2018cki,LIGOScientific:2018mvr,LIGOScientific:2020ibl,KAGRA:2021vkt} but estimates generated from witness channels such as in \textsc{GWsubtract} are also possible~\cite{Davis:2022ird}.
Glitch-subtracted data are then used for downstream source inference.
This method retains all the data and information available and is, therefore, more suitable for production analyses.
However, its efficacy hinges on the subtracted glitch estimate since the true morphology of the glitch cannot be perfectly known.
In the fair draw case, the expected glitch residual SNR is non-zero due to statistical uncertainty~\cite{Cutler:2005qq}.
In the witness channel case, the relevant transfer functions induce further systematic and/or statistical uncertainty~\cite{Payne:2022spz}.
Residual glitch power that could bias inference is therefore expected.

%%%%%%%%%%%%%%%%%%%%%%%%%%%%%%%%%%%%%
\subsubsection{Marginalization Over Glitch Realizations}\label{sec:marginalizing-over-glitches}
%%%%%%%%%%%%%%%%%%%%%%%%%%%%%%%%%%%%%

Since selecting a single glitch estimate to subtract results in residual glitch SNR, the final method is to marginalize over the glitch.
This approach is the most robust, but it is also typically more difficult to implement.
Given some parameterized glitch model $g(\phi)$, we can model the data as:

\begin{equation}
    d = n + h(\theta) + g(\phi)
\end{equation}

From this, we may extend the typical likelihood in a single detector to include the glitch:

\begin{multline}
        \ln \mathcal{L}(d | \theta, \phi) = - \frac{1}{2} \sum_k \biggr{\{} \frac{[d_k - h_k(\theta) - \phi_k(\phi)]^2}{S_{n}(f_k)} \\+ \ln(2\pi S_n(f_k))\biggr{\}}
\end{multline}
Where $k$ indexes the frequency bins being summed over, and $S_n(f_k)$ is the power spectral density in the $k$'th frequency bin.
In detectors without glitches this reduces to the standard CBC likelihood, and they combine in the usual way. 
Using this formulation, one may then sample over both $h(\theta)$ and $g(\phi)$ simultaneously.
From these samples, one may then marginalize over $\phi$ to produce CBC posteriors which reflect uncertainties in the modeling of the glitch.

We perform three glitch-marginalized analyses on GW191109.
First, using \textsc{BayesWave}, we combine the signal model with \textsc{IMRPhenomXPHM} described in Sec.~\ref{sec:fd-inference-methodology} and the sine-Gaussian glitch model described in Sec.~\ref{sec:glitch-modeling-bayeswave}.
Compared to previous relevant analyses~\cite{Chatziioannou:2021ezd,Hourihane:2022doe,Payne:2022spz} we have extended the signal model to support waveforms with spin-precession and higher-order modes.
Second, again using \textsc{BayesWave}, we combine the coherent wavelet signal model described in Sec.~\ref{sec:fd-inference-methodology} and the incoherent wavelet glitch model described in Sec.~\ref{sec:glitch-modeling-bayeswave}~\cite{Ghonge:2023ksb}.
This analysis uses a more flexible ---and thus less sensitive--- model for the GW signal; it is thus used as an additional check.
Even though \textsc{BayesWave} has the capability to also marginalize over the Gaussian noise PSD~\cite{Littenberg:2014oda,Chatziioannou:2019zvs}, we fix it for consistency with other analyses and since its effect on source inference is generally minimal~\cite{Plunkett:2022zmx}. 
Third, we implemented the physically-motivated scattered light glitch model of Sec.~\ref{sec:scattered-light-modeling} in \textsc{Bilby}.
This allows us to jointly use the slow scattering model and the \textsc{NRSur7dq4} approximant for the signal.

%%%%%%%%%%%%%%%%%%%%%%%%%%%%%%%%%%
\section{Understanding the GW191109 Inference}\label{sec:why-is-the-inferred-chieff-negative}
%%%%%%%%%%%%%%%%%%%%%%%%%%%%%%%%%%

In this section, we explore the relation between the GW191109 inference, especially the $\chieff<0$ measurement, and the glitch-affected data.  
In Fig.~\ref{fig:witness-spectrogram-overlay} we show spectrograms of the original data (without any glitch mitigation) in LLO at the time of the event.\footnote{A similar plot for the LHO data showing that the scattered light glitch does not overlap with the signal is given in App.~\ref{app:lho-glitch}.} 
Arch-like traces (multiple colors) show the glitch time-frequency tracks as predicted by a witness channel.
The light blue track corresponds to GW191109 as inferred with \textsc{NRSur7dq4} from data after the glitch was subtracted (Run~\ref{rid:subtracted-fd-fL20-fH20} in Table~\ref{tab:all-analyses-with-settings}).
The upper panel presents $16\,$s of data; scattering arches are visible leading up to the event. 
In the bottom panel, we focus on the vicinity of the signal and highlight the intersection of the signal track with visible excess power along the projected scattering tracks. 
The first is at ${\sim}24 \, \si{Hz}$ and has the expected duration and morphology of a scattering arch.
The second is at ${\sim}36 \, \si{Hz}$ and while it coincides with the glitch track predicted by the witness, the excess power duration is short and does not match the expected behavior of slow scattering. 
As noted in Ref.~\cite{Davis:2022ird}, this $36 \, \si{Hz}$ excess power is not included in the original \textsc{BayesWave} glitch reconstruction and thus not subtracted in the GWTC-3 data.

We begin by confirming and extending the results of Ref.~\cite{Davis:2022ird} with \textsc{NRSur7dq4}.
Analyzing data from each detector separately (Runs~\ref{rid:subtracted-fd-fL20} and~\ref{rid:subtracted-fd-fH20} in Table~\ref{tab:all-analyses-with-settings}) we confirm that the measurement is driven solely by LLO, which prefers $\chi_{\mathrm{eff}}<0$ at 99.6\%, compared to 20.0\% in LHO. 
Coherent analysis of both detectors (Run~\ref{rid:subtracted-fd-fL20-fH20} in Table~\ref{tab:all-analyses-with-settings}) tends to the LLO conclusion due to LLO's higher sensitivity in the relevant frequency range, shown below to be $20{-}40 \, \si{Hz}$.
Indeed, the maximum likelihood waveform from the coherent analysis accumulates $20\%$ ($8\%$) of its SNR squared in LLO (LHO) for frequencies below $40$\,Hz.
This estimate further suggests that LHO data cannot aid in determining whether the critical ${\sim}36\,$Hz excess power is part of the signal or the glitch.

Similar differences in parameter inference per detector are present for other parameters as well, notably the detector-frame total mass $M$ and luminosity distance $D_L$; see footnote~\ref{footnote3} for a discussion of the correlation between $\chieff$ and $D_L$.
For example, in individual detector analyses (Runs~\ref{rid:subtracted-fd-fL20} and~\ref{rid:subtracted-fd-fH20} in Table~\ref{tab:all-analyses-with-settings}) the detector-frame total mass is $M=133^{+14}_{-14}\, \mathrm{M}_\odot$ ($M=162^{+21}_{-20}\, \mathrm{M}_\odot$) in LLO (LHO), while the luminosity distance is $D_L=1630^{+1360}_{-850} \,\mathrm{Mpc}$ ($D_L=2760^{+2300}_{-1570} \,\mathrm{Mpc}$) in LLO (LHO).
The corresponding source-frame total mass remains the same as the increases in detector-frame mass and distance effectively ``cancel out".
Though different, these estimates are still consistent with each other within statistical uncertainties so there is no indication of a discrepancy across detectors as was the case for GW200129~\cite{Payne:2022spz}.
Moreover, these differences do not lead to diverging astrophysical interpretations like the $\chieff$ inference; we therefore focus on the latter in what follows.

\begin{figure*}
    \centering
    \includegraphics[width=\textwidth]{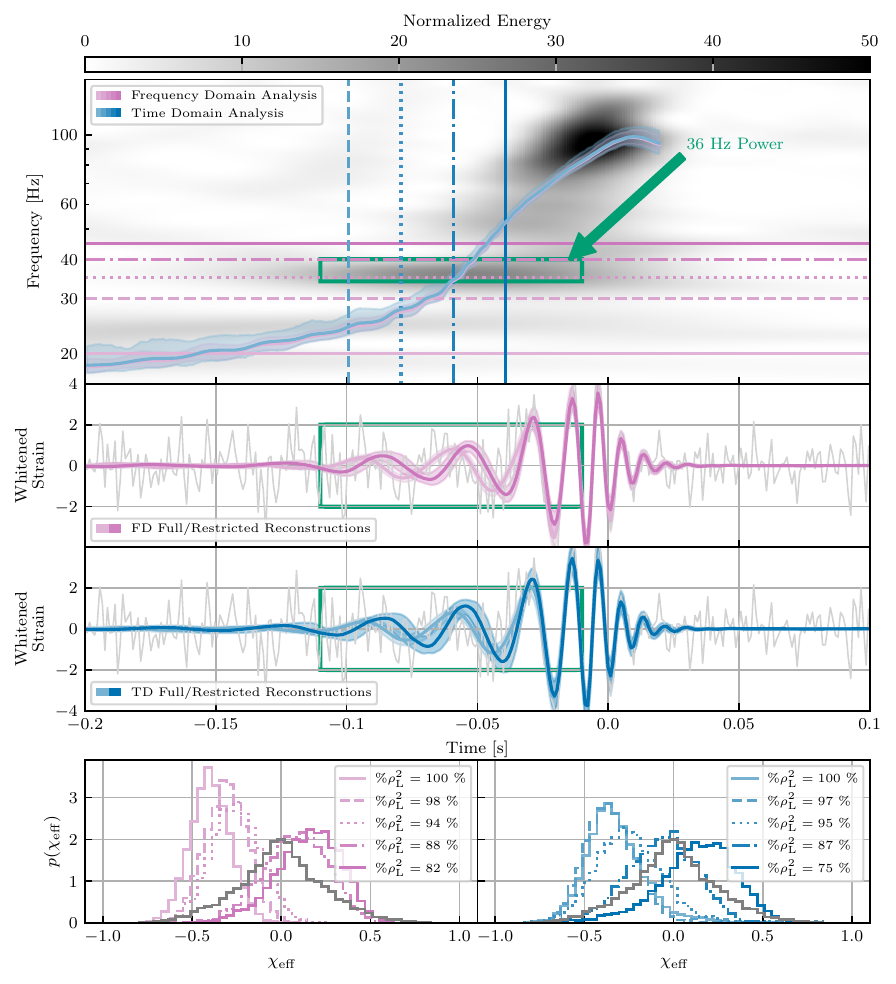}
    \caption{
    Tracing the $\chieff$ inference across frequencies and times.
    The top panel shows the spectrogram of the glitch-subtracted data around GW191109, with residual excess power at $36 \, \si{Hz}$ highlighted along with the signal track.
    We progressively remove data in the frequency domain (pink) and the time domain (blue) and reanalyze the restricted data.
    Vertical and horizontal lines in the top panel denote the time and frequency cuts, respectively; only data to the left or above these lines are analyzed.
    The two middle panels show the whitened time-domain data (grey) and signal reconstruction (pink and blue).
    Lighter colors correspond to the analyses of the full data, while darker colors correspond to the most restricted data (frequencies above $40 \, \si{Hz}$ and times from $- 0.04 \, \si{s}$ before merger onwards).
    The bottom row shows the $\chieff$ prior (gray) and $\chieff$ marginal posteriors from analyses with varying levels of data restriction, each corresponding to the lines on the top panel. 
    The legend notes the SNR squared $\rho^2$ fraction in Livingston that remains in the analysis band after each data restriction.
    }
    \label{fig:TimeFrequencyCombined}
\end{figure*}

%%%%%%%%%%%%%%%%%%%%%%%%%%%%%%%%%%%%%%
\subsection{Tracing inference across frequencies}\label{sec:fd-cuts-results}

To more precisely track the origin of the $\chieff < 0$ measurement across LLO data, we perform a series of coherent 2-detector analyses where we successively restrict the LLO frequencies, incrementing the minimum frequency $f_{L}$ by $5 \, \si{Hz}$ from $20$ to $45 \, \si{Hz}$ (Runs~\ref{rid:subtracted-fd-fL20-fH20}--\ref{rid:subtracted-fd-fL45-fH20} in Table~\ref{tab:all-analyses-with-settings}).
We use the glitch-subtracted data where the $24\,$Hz arch from Fig.~\ref{fig:witness-spectrogram-overlay} has been subtracted, but the $36\,$Hz excess power has not~\cite{KAGRA:2021vkt}.
A subset of these results are shown in Fig.~\ref{fig:TimeFrequencyCombined} (pink shading).
The top panel shows a spectrogram of the glitch-subtracted data; compared to Fig.~\ref{fig:witness-spectrogram-overlay}, there is no excess noise at ${\sim}24\,$Hz.

Marginalized posteriors for $\chieff$ are shown in the bottom panel.
The legend denotes the percentage of the total SNR squared $\rho^2$ (computed based on the maximum-likelihood full-band signal) that remains in the analysis window after each restriction. 
Removing data between $20 {-}30 \, \si{Hz}$ (solid vs dashed horizontal lines in the top panel and histograms in the bottom panel) or $30 {-}35 \, \si{Hz}$ (dashed vs dotted) removes 6\% of $\rho^2$ but does not dramatically alter inference: $\chieff<0$ is still preferred at 96.3\% for $f_L = 35 \, \si{Hz}$.
%(shifting from $\chieff=-0.36^{+0.17}_{-0.18}$ to $\chieff=-0.30^{+0.19}_{-0.20}$), 
%The posterior in $\chieff$ shifts somewhat more when we further exclude data corresponding to frequencies between $30 \, \si{Hz}$ and $35 \, \si{Hz}$ to $\chieff=-0.25^{+0.23}_{-0.22}$, but this is consistent with approaching the prior ($\chieff=0^{+0.59}_{-0.59}$).
Such small shifts are likely consistent with the SNR reduction and regression to the prior (gray).
Removing data $35 -40 \, \si{Hz}$ (dotted vs dot-dashed) removes an additional 6\% of $\rho^2$ and instead results in an abrupt shift in $\chieff$, with $\chieff<0$ now only at 32.2\%, a moderate preference for positive values.\footnote{Since the $\chieff$ prior is centered at zero, this shift to mildly positive values goes beyond regression to the prior. We attribute this to a mild $\chieff-D_L$ degeneracy that arises for merger-only signals. The uniform-in-volume prior favors larger $D_L$ and results in larger $\chieff$ to compensate for the amplitude reduction. This degeneracy is less pronounced when the signal inspiral is visible, as then $\chieff$ is constrained by the inspiral phase evolution beyond just the merger amplitude. \label{footnote3}}
Further bandwidth reduction does not modify the $\chieff$ posterior substantially (solid dark pink).
These results indicate that it is the LLO data between $35$ and $40 \, \si{Hz}$ that are crucial for measuring $\chieff$, coinciding with the $36\,$Hz excess power visible in both the original data, Fig.~\ref{fig:witness-spectrogram-overlay}, and the glitch subtracted data, Fig.~\ref{fig:TimeFrequencyCombined}.

The second panel from the top of Fig.~\ref{fig:TimeFrequencyCombined} shows the whitened time-domain reconstructions.
We compare signal reconstructions from two analyses with dramatically different $\chieff$ posteriors: the full bandwidth analysis that prefers $\chieff<0$ against the $f_L=40\,$Hz analysis with a mildly positive $\chieff$.
While the two analyses are conducted on different data subsets, we can still evaluate the waveforms across the same times and plot them together.
The reconstructions are consistent during the merger (corresponding to high frequencies included in both analyses), but start diverging $2-3$ cycles before merger.
By eye, the full-band reconstruction better matches the data for $t\approx-0.6\,$s, corresponding to the $36\,$Hz excess power.
When that power is included in the analysis, the signal model absorbs it by setting $\chieff<0$ and pushing the GW cycle to earlier times.
If that power is not part of the analysis, $\chieff$ is no longer required to be negative and the $36\,$Hz excess power is left unaccounted for.

This conclusion raises the question of whether the 36\,Hz excess power is part of the signal or part of a glitch that remained unsubtracted.
Though the shift in the $\chieff$ posterior is suggestive of anomalous noise, it is possible that it is at least partly due to loss of information as $6\%$ of $\rho^2$ in LLO is contained in the $35 {-} 40 \, \si{Hz}$ frequency band.
In Sec.~\ref{sec:injection-results} we contextualize this $\chieff$ shift with simulated signals.

%%%%%%%%%%%%%%%%%%%%%%%%%%%%%%%%%%%%%
\subsection{Tracing inference across times}\label{sec:td-cuts-results}
%%%%%%%%%%%%%%%%%%%%%%%%%%%%%%%%%%%%%

%\RU{Pending identification of the issue with TD inference, we may add discussion of how the fine grained time slices inform our understanding of power distribution over the signal (and can also reference this to the ``weird" injections)}

Having identified the crucial frequencies for $\chieff$ inference, here we do the same across time with the time-domain analysis described in Sec.~\ref{sec:td-inference-methodology}. 
When used on the full dataset, frequency- and time-domain analyses should yield equivalent results.
Indeed, we find consistent posteriors for $\chieff$ when analyzing GW191109 in the frequency and time domains, as seen by the solid histograms in Fig.~\ref{fig:TimeFrequencyCombined}.\footnote{
The time- and frequency-domain analyses employ different priors on masses, luminosity distance, and time. The time-domain inference uses priors which are uniform in detector-frame total mass, mass ratio, and luminosity distance; and are normally distributed in geocenter time, centered at $1257296855.2114642$ with a width of $0.005$ seconds. 
We confirm that the differences in time and the mass priors effect the posteriors minimally.
Reweighting between the two luminosity distance priors proves difficult due to finite sampling and upweighting portions of parameter space with no support in the posterior. 
However, the luminosity distance posteriors from the time- and frequency-domain analyses are in high agreement when the full data is analyzed, despite using different priors.}

However, as Fig.~\ref{fig:witness-spectrogram-overlay} shows, there is no 1-to-1 mapping between time and frequency for the glitch.
Though not as apparent, the same is true for the signal beyond the inspiral regime or due to spin-precession and higher-order modes.
Truncating the data in the time domain is, therefore, not equivalent to truncating in the frequency domain, as the former allows us to probe the
effect of individual cycles (or parts of cycles) of the signal or the glitch. 

Results from progressively excluding the earlier portion of the signal in the time-domain (Run~\ref{rid:subtracted-td-tLx-tHx} in Table~\ref{tab:all-analyses-with-settings}) are shown in Fig.~\ref{fig:TimeFrequencyCombined} (blue shading).
We find broadly similar results as the frequency-domain analysis: the full data yield preference for $\chieff<0$.
As the segment that contains the $36\,$Hz excess power is progressively removed (blue vertical lines in the top panel), the $\chieff$ posterior shifts to being principally positive (equivalent blue histograms in the bottom panel).
Overall, the data $0.1 {-} 0.04 \, \si{s}$ before merger are crucial for $\chieff<0$ inference.
Compared to the frequency-domain results, the shift in the $\chieff$ posterior is more gradual, likely due to the fact that the $36\,$Hz power is more concentrated in frequency, hence no time ``cut" abruptly completely excludes it.
Waveform reconstructions from the time-domain analysis (third panel from the top in Fig.~\ref{fig:TimeFrequencyCombined}) yield consistent conclusions.

%%%%%%%%%%%%%%%%%%%%%%%%%%%%%%%%%%%%%
\subsection{Simulated signals}\label{sec:injection-results}
%%%%%%%%%%%%%%%%%%%%%%%%%%%%%%%%%%%%%

\begin{figure}
    \centering
    \includegraphics[width=\columnwidth]{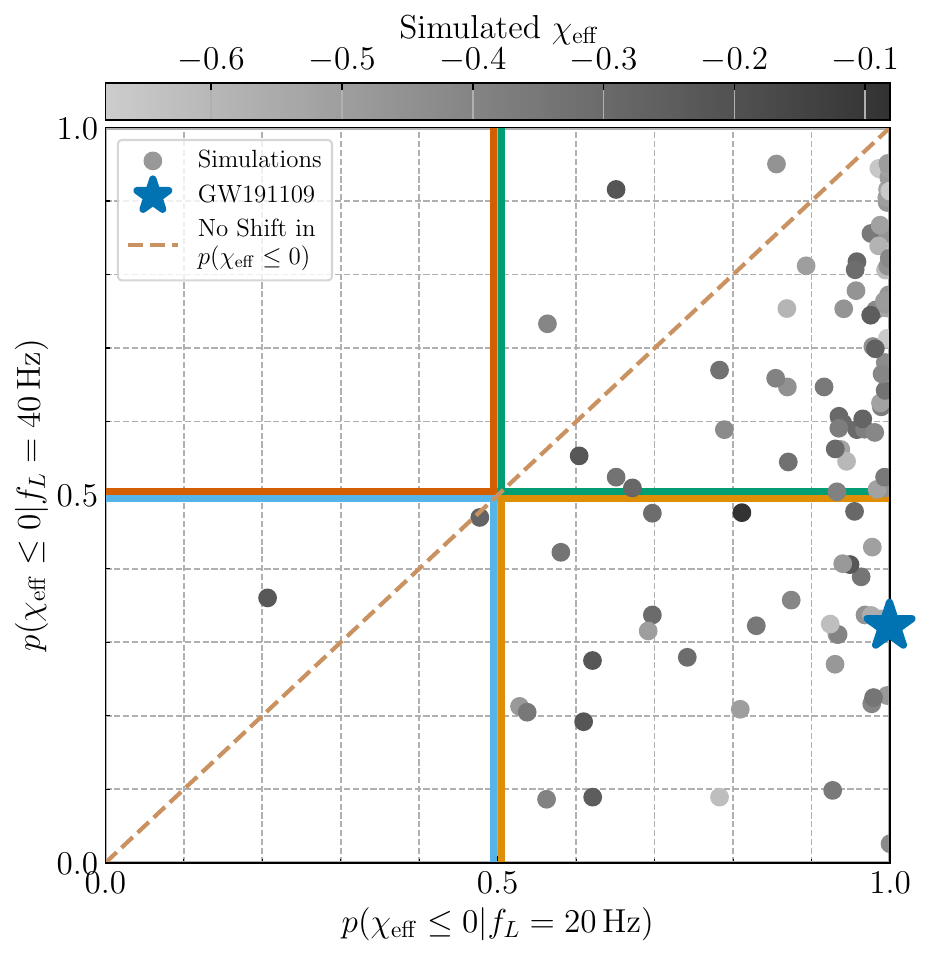}
    \caption{Shifts in the probability of $\chieff\leq0$ for $100$ simulated signals consistent with GW191109 in Gaussian noise (dots) and the real signal (cross). 
    The x-axis corresponds to a full-band analysis, while the y-axis corresponds to a restricted-band analysis with $f_L=40\,$Hz.
    Going clockwise, the top left quadrant (red-orange axes) would contain cases where the posterior shifted from majority positive to negative (of which there were none), the top right quadrant (green axes) contains cases which were majority negative in both full- and restricted-band analyses, the bottom right quadrant (orange axes) contains cases which started majority negative and became majority positive (including GW191109), and the bottom left quadrant (blue axes) contains cases which were consistently majority positive.
    The $x=y$ line (dashed brown) corresponds to no shift in the probability for $\chieff\leq 0$.
    }
    \label{fig:InjectionChiEffShifts}
\end{figure}

We investigate the degree to which the abrupt shift in the $\chieff$ posterior in Fig.~\ref{fig:TimeFrequencyCombined} is consistent with SNR loss from removing data with simulated signals.
We simulate $100$ signals drawn from the GW191109 full-band posterior (Run~\ref{rid:subtracted-fd-fL20-fH20} in Table~\ref{tab:all-analyses-with-settings}), add them to Gaussian noise drawn from the GW191109 PSDs in LLO and LHO, and analyze the full data versus the $>40\,$Hz data in LLO independently (Runs~\ref{rid:simulated-full-band-start}--\ref{rid:simulated-restricted-band-end} in Table~\ref{tab:all-analyses-with-settings}).
Signals have true values $\chieff<0$ but as data and signal SNR are removed when $f_L=40\,$Hz, we expect the posterior to become more prior-like and shift toward $\chieff=0$.
%However, subtle correlations between parameters mean that when the full posterior becomes more prior like, it is possible for the \emph{marginalized} $\chieff$ posterior to shift in ways which may defy expectations, but which would appear in a consistent manner for injections into Gaussian noise.
For each simulated signal, Fig.~\ref{fig:InjectionChiEffShifts} shows the probability of $\chieff\leq0$ from the full-data, $f_L=20\,$Hz, and the restricted-data, $f_L=40\,$Hz, analysis.

For almost all signals removing low-frequency LLO data results in a $\chieff$ posterior that shifts closer to the prior and positive values (lying below the diagonal) as expected for signals with true values of $\chieff<0$.
In most cases, this shift is marginal, and posteriors stay majority-negative, as evidenced by the high density (64\% of all signals) in the top right quadrant (green axes).
The next most likely outcome is the bottom right quadrant (orange axes), which contains 34\% of the signals, including GW191109: here the $\chieff$ posterior shifts from favoring negative to positive values.
Among these, GW191109 is one of the more extreme cases, exhibiting a shift more significant than 94\% of the simulations. 
Therefore, we conclude that the $\chieff$ shift presented in Fig.~\ref{fig:TimeFrequencyCombined} is \textit{unlikely}, but not impossible, to be explained by a random Gaussian noise instantiation, i.e.~without needing to invoke residual glitch power.
In App.~\ref{sec:chisq-methodology} we present further results based on a $\chi^2$ test used in search algorithms that tracks how SNR is accumulated along the signal~\cite{Allen:2004gu, Usman:2015kfa, Davis:2020nyf}.
Consistent with Fig.~\ref{fig:InjectionChiEffShifts}, the test is inconclusive: the full-band analysis (Run~\ref{rid:subtracted-fd-fL20-fH20} in Table~\ref{tab:all-analyses-with-settings}) has behavior more extreme than most simulations, but it is not strongly inconsistent with them.

%%%%%%%%%%%%%%%%%%%%%%%%%%%%%%%%%%
\section{Glitch-Marginalized Inference}\label{sec:application-of-glitch-mitigation}
%%%%%%%%%%%%%%%%%%%%%%%%%%%%%%%%%%

Having established that the 35--40\,Hz data drive the negative $\chieff$ inference, we turn to the question of whether these data are meaningfully impacted by residual glitch power. 
We go beyond subtracting a single estimate for the glitch and simultaneously model both the signal and the glitch as described in Sec.~\ref{sec:marginalizing-over-glitches}. 
All analyses in this section use the original data in both detectors with no prior glitch mitigation. 
While this approach is robust against residual glitch power from subtracting a single glitch estimate, it is still impacted by modeling choices, specifically both the parametrized model (physical scattering model or wavelets) and the corresponding glitch parameter priors.

Since the 36\,Hz excess power coincides in frequency with an arch predicted by the witness channel, Fig.~\ref{fig:witness-spectrogram-overlay}, it is reasonable to expect it to be part of the scattering event and thus a prime target for the slow scattering model~\cite{Udall:2022vkv}.
However, the time-frequency morphology of the 36\,Hz excess power does not resemble scattering arches, which motivates the alternative wavelet-based glitch model.
In principle, \textsc{BayesWave} can fit any excess power by adding enough wavelets.
Such a many-wavelet fit might be statistically disfavored, though, as it relies on a large number of parameters and a reduced posterior-to-prior volume.
The exact quantitative impact of this Occam penalty is controlled by the wavelet parameter priors, which influence whether it is statistically favorable to add a wavelet to capture the excess power or instead attribute it to the signal.
The most influential prior is likely the one for the wavelet amplitude, which ---although broad--- favors wavelets with SNR ${\sim}5$.
The situation is further complicated by the the low LHO sensitivity in the relevant frequencies, which weakens its contribution to the likelihood, making the discrimination between glitch and signal even more dependent on the prior shape.

%%%%%%%%%%%%%%%%%%%%%%%%%%%%%%%%%%%%%
\subsection{Slow scattering glitch model}\label{sec:glitch-reconstructions}
%%%%%%%%%%%%%%%%%%%%%%%%%%%%%%%%%%%%%

\begin{figure*}
    \centering
    \includegraphics[width=\textwidth]{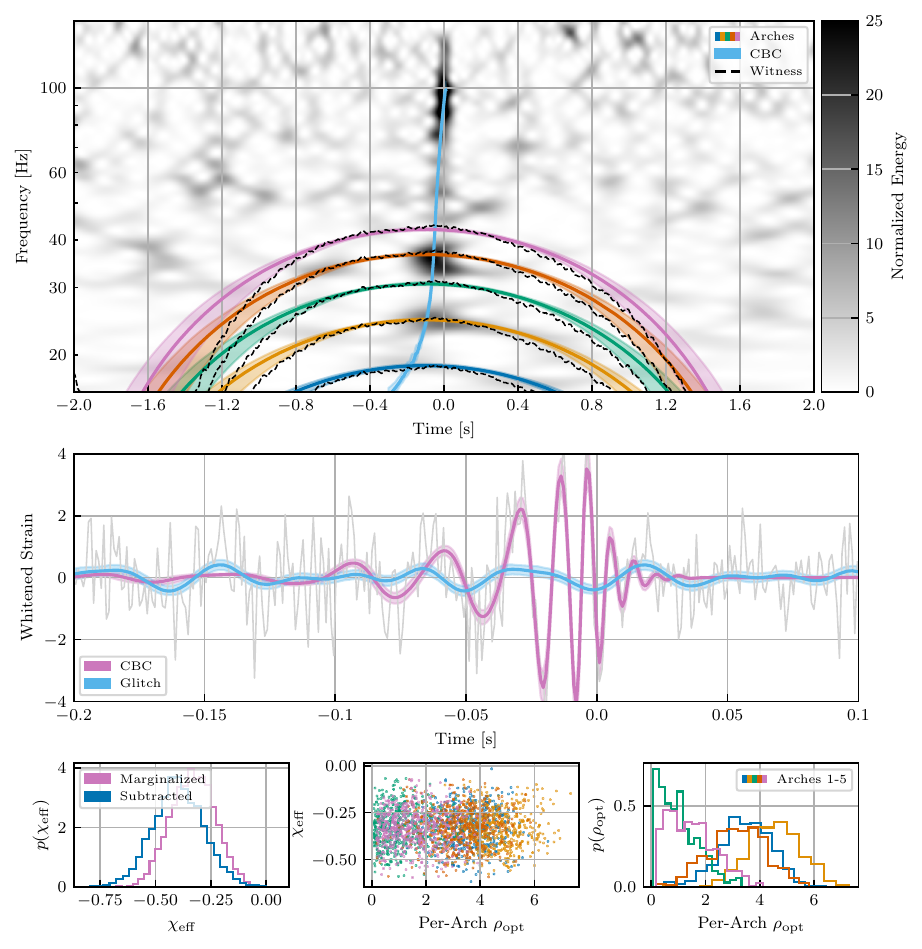}
    \caption{Jointly modeling the glitch with the physical slow scattering model and the signal with \textsc{NRSur7dq4} (Run~\ref{rid:scattering-uniform-targeted} in Table~\ref{tab:all-analyses-with-settings}).
    In the top panel, we show a spectrogram of the data, along with the posterior for the glitch arches (median and 90\% credible intervals; multiple colors), the signal track (blue), and the prediction of the witness channel (black dashed).
    In the middle panel, we show the whitened time-domain posterior reconstruction for the glitch (blue) and the signal (CBC; pink).
    In the bottom left panel, we show the marginalized $\chieff$ posterior from this analysis (pink), along with the equivalent result from glitch-subtracted data (Run~\ref{rid:subtracted-fd-fL20-fH20} in Table~\ref{tab:all-analyses-with-settings}; blue).
    In the bottom right panel, we show the marginalized posterior for the optimal SNR of each individual arch.
    Finally, in the bottom middle panel, we show a scatter plot of individual posterior samples in the $\rho_{\rm opt}-\chieff$ plane for each arch, showing that no correlation exists.
    }
    \label{fig:slow-scattering-full-plot}
\end{figure*}

We begin with the scattered light model in Fig.~\ref{fig:slow-scattering-full-plot} (Run~\ref{rid:scattering-uniform-targeted} in Table~\ref{tab:all-analyses-with-settings}), which models five arches with a uniform amplitude prior and the ``Targeted" modulation prior that is informed by the witness motion.
The signal is modeled with \textsc{NRSur7dq4}.
The top panel shows a spectrogram of the data and the signal and glitch posteriors.
The inferred glitch arches (multiple colors) match the witness prediction for the arch peak frequency spacing (${\sim}6 \, \si{Hz}$) in the region of maximum glitch power.
The optimal SNR $\rho_{\rm opt}$ posterior for each arch is shown in the bottom right panel, which reveals that three non-consecutive arches are confidently recovered with $\rho_{\rm opt}>0$: the first one at $18 \, \si{Hz}$ (blue), the second at $24 \, \si{Hz}$ (yellow), and the fourth at $36 \, \si{Hz}$ (orange).
The third arch at $30 \, \si{Hz}$ has negligible SNR, $\rho_{\rm opt}<2$ at 88\% credibility.
Though seemingly surprising given the physical interpretation of scattered light based on bounces off of moving surfaces, a varying arch amplitude is commonly observed and the SNR further depends on the noise PSD that decreases with frequency in this range.
The full glitch reconstruction in the time domain is plotted in the middle panel (blue) along with the signal (pink).
As expected from the presence of multiple arches, the glitch does not have a constant frequency.

The $\chieff$ inference is presented in the bottom row.
The bottom left panel shows the marginalized $\chieff$ posterior from this analysis (pink).
For comparison, we also plot the posterior from the standard two-step analysis where the glitch has been pre-subtracted and only the signal is analyzed (Run~\ref{rid:subtracted-fd-fL20-fH20} in Table~\ref{tab:all-analyses-with-settings}; blue).
Under glitch marginalization, $\chieff$ remains definitively negative at ${\sim} 100\%$ credibility, though the median
increases from $-0.40$ to $-0.33$.
The glitch and $\chieff$ inference are uncorrelated, as shown in the bottom middle panel through a scatter plot for $\chieff$ and the optimal SNR of each arch.\footnote{Rather than the $36\,$Hz power, we attribute the small shift in the $\chieff$ median in Fig.~\ref{fig:slow-scattering-full-plot} to the particular glitch realization that was subtracted for the GWTC-3 analysis. Indeed when we analyze the original data (no glitch mitigation) with only a signal (Run~\ref{rid:original-fd-fL20-fH20} in Table~\ref{tab:all-analyses-with-settings}), we obtain a $\chieff$ posterior more similar to that of the marginalized analysis with a median $\chieff$ of $-0.36$.}
This suggests that even though there is a glitch arch at $36\,$Hz its time-frequency morphology does not match the $36\,$Hz excess power.
Even when the signal and glitch are simultaneously modeled, most of the $36\,$Hz excess power is attributed to the signal and results in $\chieff<0$.
The time-domain reconstructions in the middle panel confirm this interpretation, with the signal reconstruction closely resembling those in Fig.~\ref{fig:TimeFrequencyCombined}, while a lower-amplitude glitch oscillation accounts for the remainder.
We have verified that these $\chieff$ results are robust under alternative, yet reasonable, priors for the glitch: log-uniform in amplitude and the ``Physical" modulation prior discussed in Sec.~\ref{sec:scattered-light-modeling} (Runs~\ref{rid:scattering-uniform-physical},~\ref{rid:scattering-loguniform-targeted}, and~\ref{rid:scattering-loguniform-physical} in Table~\ref{tab:all-analyses-with-settings} for uniform amplitude with physical modulation, log-uniform amplitude with targeted modulation, and log-uniform amplitude with physical modulation respectively). 
We have also verified that other parameters, such as the binary total mass and mass ratio, remain consistent between glitch-subtracted and glitch-marginalized analyses.

To summarize, we conclude that the $36\,$Hz power is not exclusively due to the signal.
Not only does the witness channel predict some glitch power, but also the slow scattering model places an $\rho_{\rm opt}\sim 3$ arch, notably louder than its adjacent arches. 
However, the excess power is not entirely attributed to scattered light as it is morphologically inconsistent with a slow scattering arch.\footnote{In App.~\ref{sec:appendix-effect-of-glitch-priors} we show that unphysical priors on the slow scattering parameters can indeed twist the model into fully absorbing the $36\,$Hz power and eliminating the $\chieff<0$ inference. Such priors are, however inconsistent with slow scattering, which forms the basis of the glitch model to being with.} 
The $\chieff<0$ inference, therefore, persists under the physical slow scattering interpretation of this glitch.

%%%%%%%%%%%%%%%%%%%%%%%%%%%%%%%%%%%%%
\subsection{Wavelet glitch model}\label{sec:bayeswave-glitch-marginalized-results}
%%%%%%%%%%%%%%%%%%%%%%%%%%%%%%%%%%%%%

\begin{figure*}
    \centering
    \includegraphics[width=\textwidth]{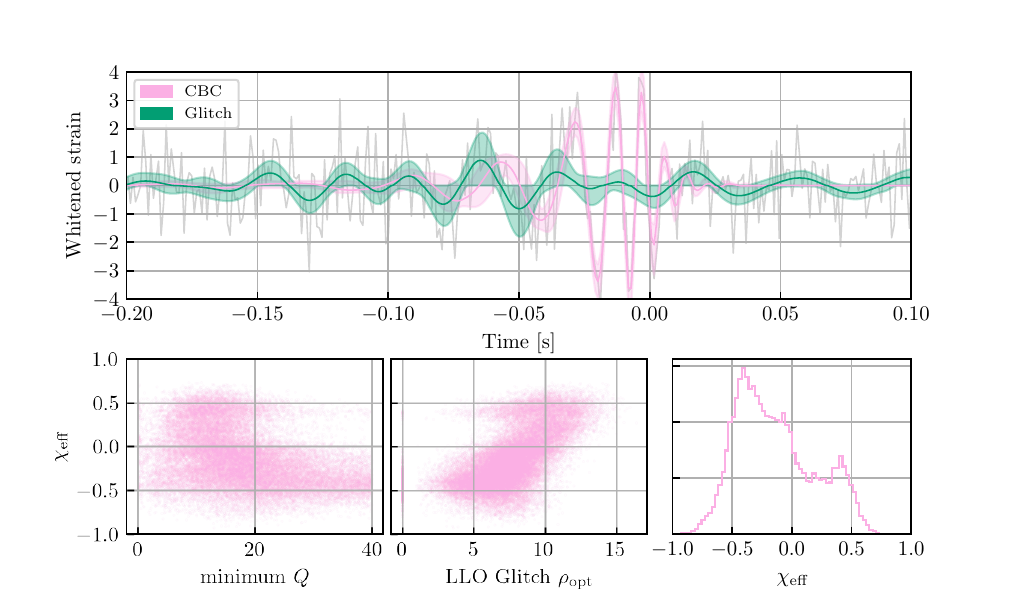}
    \caption{
    Jointly modeling the glitch with sine-Gaussian wavelets and the signal with \textsc{IMRPhenomXPHM} (Run~\ref{rid:bw-cbc-plus-glitch} in Table~\ref{tab:all-analyses-with-settings}). 
    The top panel shows the whitened time-domain data (grey) and median and $90\%$ credible intervals for the glitch (green) and signal (CBC; pink). 
    The bottom row displays marginalized posteriors. 
    The right panel shows the glitch-marginalized $\chieff$ posterior, which displays a much larger spread than the results of Fig.~\ref{fig:slow-scattering-full-plot}, now being consistent with $\chieff=0$. 
    The left panel shows the scatter plot between $\chieff$ and the minimum quality factor $Q$ among all wavelets of each posterior sample.
    Positive $\chieff$ is correlated with low $Q$.
    Scattered light is characterized by larger $Q$-values~\cite{Hourihane:2022doe}, confirming that $\chieff>0$ only if the glitch does not match the expected scattered light morphology. 
    The middle panel shows a scatter plot between $\chieff$ and the glitch SNR which are again correlated: higher glitch power leads to a more positive $\chieff$. 
    %\kc{Remember to update the $\chieff$ plot during review to add more ticks on the x-axis and make it symmetric}
    }
    \label{fig:bayeswave-marginalized-posterior}
\end{figure*}

\begin{figure*}
    \centering
    \includegraphics[width=\textwidth]{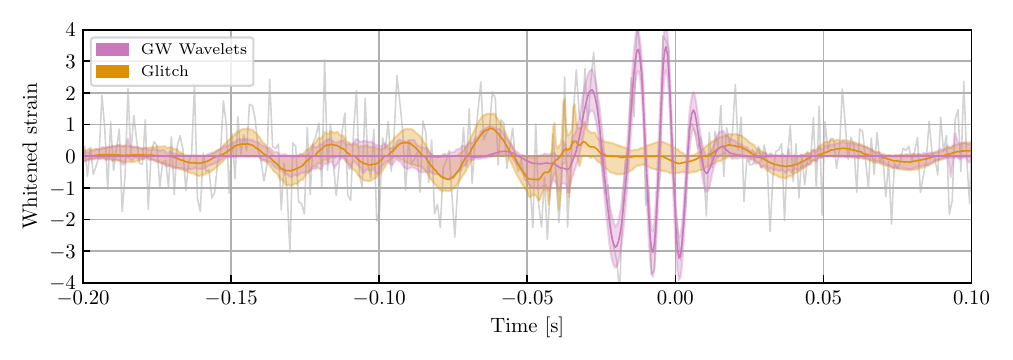}
    \caption{Jointly modeling both the glitch and the signal with sine-Gaussian wavelets (Run~\ref{rid:bw-gwwavelets} in Table~\ref{tab:all-analyses-with-settings}).
    We plot the whitened time-domain data (grey) and median and $90\%$ credible intervals for the glitch (orange) and signal (GW Wavelets; purple).
    The $36\,$Hz excess power is consistent with originating from either the glitch or the signal at the 90\% credible level.
    }
    \label{fig:bayeswave-gwWavelets}
\end{figure*}

The physically-motivated slow scattering model finds some glitch power at $36\,$Hz but cannot account for the entire $36\,$Hz excess power.
This might be because of modeling systematics, the presence of other (beyond slow-scattering) non-Gaussian noise, or simply because the $36\,$Hz excess power is indeed part of the signal. 
We explore these possibilities with \textsc{BayesWave} and its more flexible wavelet-based glitch model as described in Sec.~\ref{sec:glitch-modeling-bayeswave}. 
We present two analyses: both marginalize over the glitch with wavelets but the GW signal is modeled with either the compact binary model \textsc{IMRPhenomXPHM} or with coherent wavelets. 

%%%%%%%%%%%%%%%%%%%%%%%%%%%%%%%%%%%%%
\subsubsection{\textsc{IMRPhenomXPHM}}

In Fig.~\ref{fig:bayeswave-marginalized-posterior} we show results from the joint analysis with \textsc{IMRPhenomXPHM} for the signal and wavelets for the glitch (Run~\ref{rid:bw-cbc-plus-glitch} in Table~\ref{tab:all-analyses-with-settings}). 
The top panel shows the whitened time-domain reconstructions.
Compared to the reconstructions in Fig.~\ref{fig:slow-scattering-full-plot} there is now increased uncertainty around the $36\,$Hz excess power, i.e. between times $-0.09$ and $-0.04\,$s. 
This is due to the larger flexibility of the glitch model, which can now compete with the signal for the data around $-0.06\,$s, leading to larger uncertainties for both models.
The larger uncertainty is also reflected in the glitch-marginalized $\chieff$ posterior shown in the bottom right panel. 
Compared to Fig.~\ref{fig:slow-scattering-full-plot}, the $\chieff$ posterior is now much wider and entirely consistent with zero.
It displayes a broadly bimodal structure with one mode favoring $\chieff<0$ and peaking at ${\sim} -0.4$ and the other favoring $\chieff>0$ and peaking at at ${\sim} 0.4$. 
The antialigned mode is weakly favored at $70\%$ of the posterior samples have $\chieff<0$.

The increased $\chieff$ uncertainty is entirely due to the glitch and the competition between the signal and the glitch models.
The bottom middle panel shows a  posterior scatter plot for $\chieff$ and the SNR of the glitch in LIGO Livingston.\footnote{This analysis allows for glitches in both detectors, but the Hanford data are consistent with no glitch power in the analysis window.} 
The glitch SNR is strongly correlated with $\chieff$: a higher glitch power results in a more positive $\chieff$.
A small fraction of posterior samples, $\sim6\%$, have vanishing glitch SNR (zero wavelets) and a strongly negative $\chieff$, consistent with results from Fig.~\ref{fig:slow-scattering-full-plot}.
Besides the glitch power, we examine the recovered glitch morphology in the bottom left panel, where we plot $\chieff$ against the minimum quality factor among wavelets in a particular posterior sample.
The quality factor corresponds to the number of cycles in a wavelet, therefore scattering arches are characterized by larger values of $Q$~\cite{Hourihane:2022doe}. 
This plot confirms the conclusions of Fig.~\ref{fig:slow-scattering-full-plot}: if the glitch is scattering-like (large $Q$), the model cannot capture the $36\,$Hz power, and $\chieff$ tends to be negative. 
Support for $\chieff>0$ requires low values of $Q$ which morphologically do not resemble scattering arches.

These results are qualitatively robust against different glitch priors. 
%\sh{TODO, reanalyze with converged results when Katerina's finish}
When using a prior for the amplitude of each wavelet that peaks at an SNR of 3 (instead of the default value of 5), we recover the same bimodal solution for $\chieff$ and the glitch SNR.
However, the preference for the antialigned mode shifts from $70\%$ to $60\%$ suggesting that our quantitative results are impacted by the glitch prior at the few percent level.
This shift is attributed to the fact that the updated prior makes it easier to low-SNR wavelets to be added to the posterior and thus capture the $36\,$Hz excess power away from the signal model.
The impact of glitch priors is akin to the impact of compact-binary parameters on inference~\cite{Vitale:2017cfs} and is expected to be more prominent for low-SNR glitches.

We perform a final sanity check by comparing the total (signal plus glitch) reconstructions of posterior samples with $\chieff > 0$ to those with $\chieff < 0$. 
Although the two posterior modes result in different interpretations of which parts of the data are signal and which are glitch, their sums are consistent with each other.
This is expected as it is the \emph{total} strain of signal-plus-glitch that is compared to the data to calculate the likelihood. 
So any solution must result in the same total strain.
While we view this as a sanity check on the analysis convergence, it also suggests that there are $2$ distinct ways to model the data, and this analysis does not strongly prefer one over the other.

%%%%%%%%%%%%%%%%%%%%%%%%%%%%%%%%%%%
\subsubsection{Coherent wavelet model}\label{sec:glitch-marginalized-results}

\begin{figure*}
    \centering
    \includegraphics[width=\textwidth]{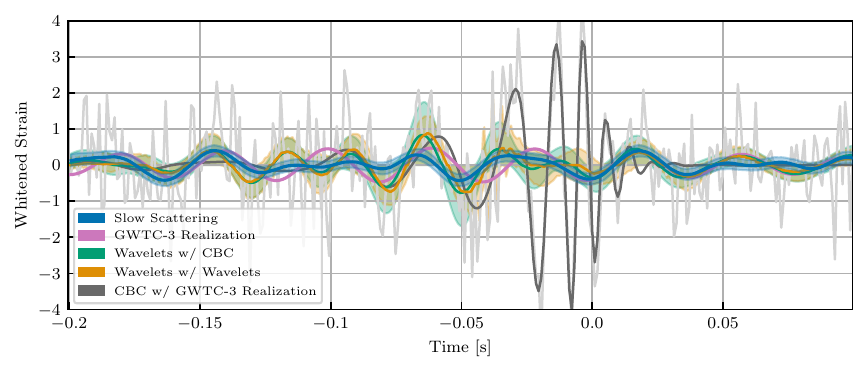}
    \caption{Comparison of reconstructions for the LIGO Livingston glitch that overlapped with GW191109 obtained by various analyses.
    The data are shown in grey, and for reference, we also show the maximum-likelihood GW reconstruction from the full-band analysis on the glitch-subtracted data in black (Run~\ref{rid:subtracted-fd-fL20-fH20} in Table~\ref{tab:all-analyses-with-settings}).
    The single realization subtracted for the GWTC-3 analysis is shown in pink~\cite{KAGRA:2021vkt}.
    The glitch inferred from the joint slow scattering and \textsc{NRSur7dq4} analysis (Run~\ref{rid:scattering-uniform-targeted} in Table~\ref{tab:all-analyses-with-settings}) is shown in blue.
    The glitch inferred with wavelets is shown in green when the signal is modeled with \textsc{IMRPhenomXPHM} (Run~\ref{rid:bw-cbc-plus-glitch} in Table~\ref{tab:all-analyses-with-settings}) and orange when the signal is also modeled with wavelets (Run~\ref{rid:bw-gwwavelets} in Table~\ref{tab:all-analyses-with-settings}).
    }
    \label{fig:compare-glitch-reconstructions}
\end{figure*}

For completeness, we present a final analysis where both the glitch and the GW are modeled with sums of wavelets~\cite{Ghonge:2023ksb} (Run~\ref{rid:bw-gwwavelets} in Table~\ref{tab:all-analyses-with-settings}). 
Since the signal model is now also phenomenological, we do not extract any binary parameters such as $\chieff$ which has thus far been guiding our conclusions. 
Instead, we directly interpret the time-domain reconstructions in Fig.~\ref{fig:bayeswave-gwWavelets}. 
As expected, using more flexible models results in increased uncertainties. 
The $36\,$Hz ($-0.06$\,s in the plot) power is still traded between the two models, and neither can rule out that it belongs to them at the 90\% credible level. 
In contrast to the signal reconstructions thus far, Figs.~\ref{fig:slow-scattering-full-plot} and~\ref{fig:bayeswave-marginalized-posterior}, the coherent wavelet model is not able to confidently recover the signal inspiral between times $-0.1$ and $-0.04$\,s. 
This is again due to the large flexibility of the wavelet signal model, which needs to extract each portion of the signal independently of the others~\cite{Ghonge:2020suv} as opposed to the waveform model that coherently models the whole signal across inspiral and merger. 

%%%%%%%%%%%%%%%%%%%%%%%%%%%%%%%%%%%%%%%%%
\subsection{Comparing glitch reconstructions}\label{sec:glitch-comparison}
%%%%%%%%%%%%%%%%%%%%%%%%%%%%%%%%%%%%%%%%%

Finally, we compare glitch reconstructions from the various glitch inferences considered in Fig.~\ref{fig:compare-glitch-reconstructions}.
The comparison includes the single glitch realization considered in GWTC-3~\cite{KAGRA:2021vkt} and the three glitch-marginalized analyses presented in this study, Figs.~\ref{fig:slow-scattering-full-plot},~\ref{fig:bayeswave-marginalized-posterior}, and~\ref{fig:bayeswave-gwWavelets}.
The glitch reconstructions are largely consistent with each other, with the largest differences encountered in the crucial $-0.06\,$s region.
As expected, the wavelet-based reconstructions have a larger statistical uncertainty due to the larger model flexibility.
This allows them to reach a larger amplitude at $-0.06\,$s which is necessary in order to capture the $36\,$Hz excess power.

%%%%%%%%%%%%%%%%%%%%%%%%%%%%%%%%%%
\section{Conclusions}\label{sec:conclusion}
%%%%%%%%%%%%%%%%%%%%%%%%%%%%%%%%%

When seeking to interpret GW data in the presence of glitches, absolute confidence in all aspects of the analysis is impossible.
Unlike compact-binary signals for which we have exact numerical relativity simulations to compare models against, glitch modeling does not have the luxury of a ``ground truth" solution.  
Nonetheless, we have sought an understanding of GW191109, its astrophysically-influential $\chieff<0$ inference, and the overlapping glitch within the limitations of imperfect glitch models and large statistical uncertainties. 

We showed that the $\chieff<0$ measurement is attributed to a segment of LIGO Livingston data occurring between $0.1$ and $0.04\,$s before the merger, and between $30$ and $40 \, \si{Hz}$.
These data are impacted by excess scattered light non-Gaussian noise, consistent with Ref.~\cite{Davis:2022ird}.
%This observation aligns with the special role played by the late inspiral and merger phases in the inference of spins.
%Furthermore, we demonstrated with simulations that it is improbable but not impossible to see transitions of this type when considering sources in this region of parameter space and the effects of Gaussian noise. 
%Limited band analyses such as these are an important tool in understanding the intricacies of parameter inference for gravitational waves, but require careful understanding of the properties of the Gaussian noise and of the signal manifold before conclusions may be drawn about the potential impact of glitches. 
%Having identified the data which is responsible for the interesting qualities of this event, we then compare the traditional form of glitch mitigation - in which CBC only analysis is performed on data from which a single glitch realization has been subtracted - to two analyses which marginalize over glitch realizations.
%In doing so, we also explore the difference between a model which is physically motivated and narrowly defined, and a model which is more flexible and generic.
Simultaneously modeling the GW signal with compact-binary waveforms and the glitch yields results that depend on the glitch model.
A physical glitch model tailored to slow scattering glitches cannot morphologically match the excess power observed in the $36\,$Hz range. 
Therefore the $\chieff<0$ measurement still stands.
A more flexible wavelet-based glitch model is instead able to fully account for the $36\,$Hz excess power and wipe out all support for $\chieff<0$.
Though witness channel information suggests that slow scattering was indeed what occurred during GW191109, we cannot rule out shortcomings of the slow scattering parametrized model or additional non-Gaussian noise.

%We find that this actually leads to substantially different conclusions, with the assumptions about the nature of the glitch determining whether the $36 \, \si{Hz}$ power is attributed to the glitch or to the CBC.
%Interestingly, we find that results with the physically motivated model vary relatively little from those produced with a single \textsc{BayesWave} realization, whereas the marginalized \textsc{BayesWave} results show much greater changes.

Given this, we cannot make absolute statements about the properties of GW191109.
If, as expected from witness channel information, the data contain Gaussian noise, a well-modeled slow scattering glitch, and a GW signal, then GW191109 likely had asymmetric masses and $\chieff<0$, strongly implying a dynamical origin~\cite{Zhang:2023fpp}.
However, if other non-Gaussian noise was present in the data, or the glitch morphology varied from classical slow scattering, spin inference becomes uninformative --- though in any situation, GW191109 remains one of the heaviest observations to date.
Distinguishing between these interpretations is challenging.
Firstly, LIGO Hanford's sensitivity in the relevant frequency range is diminished, it can therefore not contribute to the question of whether the crucial $36\,$Hz power is coherent (and thus part of the signal) or incoherent (and thus part of the glitch).
Secondly, the overall low SNR of the glitch makes results depend on the glitch model priors, e.g. the \textsc{BayesWave} glitch prior explored in Sec.~\ref{sec:bayeswave-glitch-marginalized-results}.  

%Some of these prior assumptions are implemented in the analyses themselves, with priors placed on glitch parameters in \textsc{BayesWave} being notably impactful.
%Others are prior assumptions about the nature of the glitch, namely how improbable it would be to have significant deviations from the typical slow scattering morphology,  or a second glitch coincidentally placed in time and frequency.

Our analysis builds upon Refs.~\cite{Payne:2022spz, Davis:2022ird} to propose a framework for in-depth analyses of glitch-afflicted data. 
The framework includes cross-detector comparisons, band- and time-limited analyses, simulated signals, marginalizing over the glitch, and exploring different glitch models (tailored to a specific glitch family or flexible) and prior assumptions.

As GW astronomy collects more data and seeks to constrain increasingly more subtle effects, mitigating systematics related to data quality presents a complementary challenge to waveform systematics.
Similar to waveform systematics, data quality systematics can be particularly troublesome for spin inference, which typically leaves a subtle imprint on the data and is concentrated on a small (time or frequency) region of data.
Studies such as the ones presented here and in Ref.~\cite{Payne:2022spz} are based on targeted, intensive follow-up of selected events, hand-chosen for the astrophysically important inference.
Data quality systematics aggregating over catalogs of detections require additional care to identify and mitigate in an automated way, e.g.,~\cite{Heinzel:2023vkq}.
Such efforts will be significantly aided by the work of experts in reducing the absolute rate of glitches, in characterizing the state of the detectors, and in developing efficient and statistically sound analyses in the presence of glitches.
In this work we present techniques to help address these challenges moving forward.

%%%%%%%%%%%%%%%%%%%%%%%%%%%%%%%%%%%%%
\acknowledgments
%%%%%%%%%%%%%%%%%%%%%%%%%%%%%%%%%%%%%

We thank Jess McIver, Niko Lecoeuche, Eric Thrane, Paul Lasky, Hui Tong, Maya Fishbach, Ethan Payne, and Jacob Golomb for helpful discussions about this analysis.
We thank Lucy Thomas and Aaron Zimmerman for insightful discussions on the behavior of merger dominated waveforms and the $\chieff-D_L$ degeneracy.
We thank Isobel Romero-Shaw and Tousif Islam for assistance with accessing and interpreting previous analyses.
We thank Colm Talbot for assistance in the implementation of glitch inference in \textsc{bilby}.
We thank Carl-Johan Haster for helpful comments on the manuscript.

RPU and DD are supported by NSF Grant PHY-2309200.
SM, SH, HD, and KC were supported by NSF Grant PHY-2110111 and NSF Grant PHY-2308770.
SH was supported by the National Science Foundation Graduate Research Fellowship under Grant DGE-1745301. 
This work was partially supported by the generosity of
Eric and Wendy Schmidt by recommendation of the Schmidt Futures program.
The Flatiron Institute is funded by the Simons Foundation.

This material is based upon work supported by NSF’s LIGO Laboratory 
which is a major facility fully funded by the 
National Science Foundation.
LIGO was constructed by the California Institute of Technology 
and Massachusetts Institute of Technology with funding from 
the National Science Foundation, 
and operates under cooperative agreement PHY-2309200. 
The authors are grateful
for computational resources provided by the LIGO Laboratory and supported by National Science Foundation
Grants PHY-0757058 and PHY-0823459.

% This research has made use of data or software obtained from the Gravitational Wave Open Science Center (gwosc.org), a service of the LIGO Scientific Collaboration, the Virgo Collaboration, and KAGRA. This material is based upon work supported by NSF's LIGO Laboratory which is a major facility fully funded by the National Science Foundation, as well as the Science and Technology Facilities Council (STFC) of the United Kingdom, the Max-Planck-Society (MPS), and the State of Niedersachsen/Germany for support of the construction of Advanced LIGO and construction and operation of the GEO600 detector. Additional support for Advanced LIGO was provided by the Australian Research Council. Virgo is funded, through the European Gravitational Observatory (EGO), by the French Centre National de Recherche Scientifique (CNRS), the Italian Istituto Nazionale di Fisica Nucleare (INFN) and the Dutch Nikhef, with contributions by institutions from Belgium, Germany, Greece, Hungary, Ireland, Japan, Monaco, Poland, Portugal, Spain. KAGRA is supported by Ministry of Education, Culture, Sports, Science and Technology (MEXT), Japan Society for the Promotion of Science (JSPS) in Japan; National Research Foundation (NRF) and Ministry of Science and ICT (MSIT) in Korea; Academia Sinica (AS) and National Science and Technology Council (NSTC) in Taiwan.
% LIGO was constructed by the California Institute of Technology 
% and Massachusetts Institute of Technology with funding from 
% the National Science Foundation, 
% and operates under cooperative agreement PHY-2309200. 

This work made use of \textsc{Numpy}~\cite{harris2020array}, \textsc{Scipy}~\cite{2020SciPy-NMeth}, \textsc{Matplotlib}~\cite{Hunter:2007}, \textsc{Lalsuite}~\cite{lalsuite}, \textsc{Dynesty}~\cite{Speagle:2019ivv}, \textsc{Gwpy}~\cite{gwpy}, \textsc{Astropy}~\cite{2022ApJ...935..167A}, \textsc{Bilby}~\cite{bilby_paper}, \textsc{Bilby\_Pipe}~\cite{bilby_pipe_paper}, \textsc{Pycbc}~\cite{alex_nitz_2024_10473621}, \textsc{Gwdetchar}~\cite{duncan_macleod_2024_12786150}, and \textsc{BayesWave}~\cite{Cornish:2014kda,Littenberg:2014oda, Cornish:2020dwh}.

\appendix

%%%%%%%%%%%%%%%%%%%%%%%%%%%%%%%%%%%
\section{Scattered light glitches in LHO}\label{app:lho-glitch}

\begin{figure*}
    \centering
    \includegraphics[width=\textwidth]{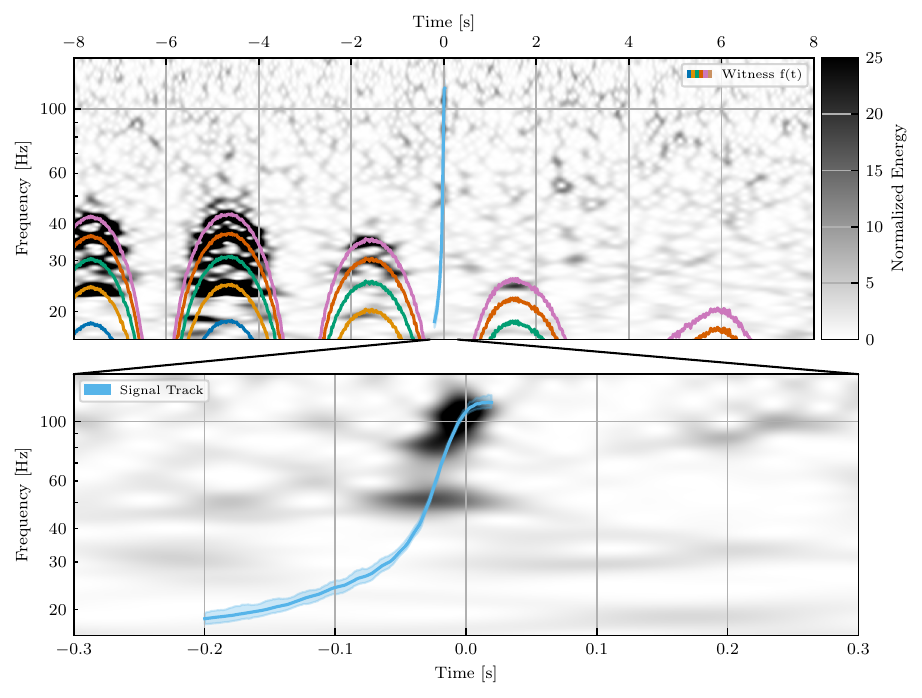}
    \caption{
    Similar to Fig.~\ref{fig:witness-spectrogram-overlay} but for LHO data at the time of GW191109, with the scattering tracks predicted by the motion of the witness channel \texttt{H1:SUS-ETMX\_L2\_WIT\_L\_DQ}.
    % \RU{Will also require waiver for this witness.}
    The absolute intensity of the slow scattering was significantly worse than in LLO, but the signal occurred at a minimum in the scattering, such that there is no overlap in time and frequency between the glitch arches and the GW191109 track (blue).
    }
    \label{fig:hanford-witness-spectrogram-overlay}
\end{figure*}

In Fig.~\ref{fig:hanford-witness-spectrogram-overlay} we show a spectrogram of the data in LHO at the time of GW191109 and the scattering tracks predicted by the witness channel.
When GW191109 entered the LHO frequency band, the scattering surface motion was at a minimum, so that the signal and the glitch are disjoint in time and frequency.
Accordingly, we expect source inference to be unaffected by the glitch.
Reference~\cite{Davis:2022ird} reached similar conclusions.
We confirm this expectation by performing analyses which restrict the frequency band in LHO in a similar fashion to main-text LLO analyses.
When restricting to $> 40 \, \si{Hz}$ in LHO (Run~\ref{rid:subtracted-fd-fL20-fH40} in Table~\ref{tab:all-analyses-with-settings}) but with no LLO restrictions, the $\chieff$ posterior remains almost entirely negative ($\chieff<0$ at 99.9\% credibility, the same as Run~\ref{rid:subtracted-fd-fL20-fH20} in Table~\ref{tab:all-analyses-with-settings} that uses all data in both detectors).
When removing sub-40\,Hz data in both detectors (Run~\ref{rid:subtracted-fd-fL40-fH40} in Table~\ref{tab:all-analyses-with-settings}), we obtain a modestly positive result ($\chieff<0$ at 33.3\% credibility), but no more so than when we only restricting the LLO data (32.2\%) (Run~\ref{rid:subtracted-fd-fL40-fH20} in Table~\ref{tab:all-analyses-with-settings}).
While it is mildly surprising that removing so much low-frequency data in LHO has so little apparent effect on inference, we attribute this to the significant difference between LLO and LHO low-frequency sensitivity.

%%%%%%%%%%%%%%%%%%%%%%%%%%%%%%%%%%%%
\section{Detailed analysis settings}
\label{app:settings}
%%%%%%%%%%%%%%%%%%%%%%%%%%%%%%%%%%%%%%%%%

\begin{table*}[]
    \begin{tabular}{|c|c|c|c|c|c|}
        \hline
        Run ID & Data & Glitch Model & Signal Model & Data Restrictions & Analysis Type \\ \hhline{|=|=|=|=|=|=|}
        \ridref{rid:subtracted-fd-fL20-fH20} & Subtracted & \longdash & \textsc{NRSur7dq4} & \longdash & \textsc{Bilby}-FD \\ \hline
        \ridref{rid:subtracted-fd-fL30-fH20} & Subtracted & \longdash & \textsc{NRSur7dq4} & $f_L = 30 \, \si{Hz}$& \textsc{Bilby}-FD \\ \hline
        \ridref{rid:subtracted-fd-fL35-fH20} & Subtracted & \longdash & \textsc{NRSur7dq4} & $f_L = 35 \, \si{Hz}$& \textsc{Bilby}-FD \\ \hline
        \ridref{rid:subtracted-fd-fL40-fH20} & Subtracted & \longdash & \textsc{NRSur7dq4} & $f_L = 40 \, \si{Hz}$ & \textsc{Bilby}-FD \\ \hline
        \ridref{rid:subtracted-fd-fL45-fH20}& Subtracted & \longdash & \textsc{NRSur7dq4} & $f_L = 45 \, \si{Hz}$ & \textsc{Bilby}-FD \\ \hline
        \ridref{rid:subtracted-fd-fL20-fH40} & Subtracted & \longdash & \textsc{NRSur7dq4} & $f_H = 40 \, \si{Hz}$ & \textsc{Bilby}-FD \\ \hline
        \ridref{rid:subtracted-fd-fL40-fH40} & Subtracted & \longdash & \textsc{NRSur7dq4} & $f_L = f_H = 40 \, \si{Hz}$ & \textsc{Bilby}-FD \\ \hline
        \ridref{rid:subtracted-fd-fL20} & Subtracted & \longdash & \textsc{NRSur7dq4} & No LHO & \textsc{Bilby}-FD \\ \hline
        \ridref{rid:subtracted-fd-fH20} & Subtracted & \longdash & \textsc{NRSur7dq4} & No LLO & \textsc{Bilby}-FD \\ \hline
        \ridref{rid:subtracted-td-tLx-tHx} & Subtracted & \longdash & \textsc{NRSur7dq4} & Various $t_H, t_L$ & TD \\ \hline
        \ridref{rid:subtracted-bilby-imrx} & Subtracted & \longdash & \textsc{IMRPhenomXPHM} & \longdash & \textsc{Bilby}-FD \\ \hline
        \ridref{rid:original-fd-fL20-fH20} & Original & \longdash & \textsc{NRSur7dq4} & \longdash & \textsc{Bilby}-FD \\ \hline
        \ridref{rid:scattering-uniform-targeted} & Original & Slow Scattering (Uniform + Targeted) & \textsc{NRSur7dq4} & \longdash & \textsc{Bilby}-FD \\ \hline
        \ridref{rid:scattering-uniform-physical} & Original & Slow Scattering (Uniform + Physical) & \textsc{NRSur7dq4} & \longdash& \textsc{Bilby}-FD \\ \hline
        \ridref{rid:scattering-loguniform-targeted} & Original & Slow Scattering (Log-Uniform + Targeted)& \textsc{NRSur7dq4} & \longdash & \textsc{Bilby}-FD \\ \hline
        \ridref{rid:scattering-loguniform-physical} & Original & Slow Scattering (Log Uniform + Physical) & \textsc{NRSur7dq4} & \longdash & \textsc{Bilby}-FD \\ \hline
        \ridref{rid:scattering-uniform-targeted-n4} & Original & Slow Scattering (Uniform + Targeted, N=4) & \textsc{NRSur7dq4} & \longdash & \textsc{Bilby}-FD \\ \hline
        \ridref{rid:scattering-uniform-unphysical} & Original & Slow Scattering (Uniform + Unphysical) & \textsc{NRSur7dq4} & \longdash & \textsc{Bilby}-FD \\ \hline
        \ridref{rid:scattering-imrphenomxphm} & Original & Slow Scattering (Uniform + Targeted) & \textsc{IMRPhenomXPHM} & \longdash & \textsc{Bilby}-FD \\ \hline
        \ridref{rid:bw-gwwavelets} & Original & Wavelets & Wavelets & \longdash & \textsc{BayesWave}-FD \\ \hline
        \ridref{rid:bw-cbc-plus-glitch} & Original & Wavelets & \textsc{IMRPhenomXPHM} & \longdash & \textsc{BayesWave}-FD \\ \hline
        \ridref{rid:simulated-full-band-start}\addtocounter{RunIDCounter}{98}-\ridref{rid:simulated-full-band-end} & Simulated & \longdash & \textsc{NRSur7dq4} & \longdash & \textsc{Bilby}-FD \\ \hline
        \ridref{rid:simulated-restricted-band-start}\addtocounter{RunIDCounter}{98}-\ridref{rid:simulated-restricted-band-end} & Simulated & \longdash & \textsc{NRSur7dq4} & $f_L = 40 \, \si{Hz}$& \textsc{Bilby}-FD \\ \hline
         
    \end{tabular}
    \caption{Settings and properties for all analyses presented in this work. 
    We list from left to right: a unique run ID hyperlinked in the text, the type of data used (original or glitch-subtracted GWTC-3 data \cite{KAGRA:2023pio, ligo_scientific_collaboration_and_virgo_2021_5546680}), how the glitch is modeled per Sec.~\ref{sec:glitch-modeling}, how the CBC signal is modeled per Sec.~\ref{sec:modeling-the-compact-binary}, frequency or time cuts on the data on top of the default settings, and the analysis type (software and data domain - \textsc{FD} for frequency and \textsc{TD} for time). 
    Analyses based on glitch-subtracted data use the data provided by GWTC-3~\cite{KAGRA:2021vkt}, while analyses that marginalize over the glitch employ the original unmitigated data.
    Frequency bands are described by $f_H$ and $f_L$ designating the minimum frequency of analysis in LHO and LLO respectively. 
    For runs which us the parameterized slow scattering model, the parenthetical descriptions correspond to the choice of amplitude prior and modulation frequency prior respectively for each run.
    All slow scattering analyses model five slow scattering arches, with the exception of Run~\ref{rid:scattering-uniform-targeted-n4}.
    }
    \label{tab:all-analyses-with-settings}
\end{table*}

In this Appendix we provide details about the settings of all analyses presented in this study. 
Table~\ref{tab:all-analyses-with-settings} identifies all analyses with a unique index, referenced throughout the text. 
We also list the data analyzed, the relevant glitch and signal models, any restrictions applied to the data being analyzed, and the analysis type (both the software used and the data domain in which it operates).
Data for these analyses is made public in the associated zenodo dataset \cite{udall_2025_14611118}.

%%%%%%%%%%%%%%%%%%%%%%%%%%%%%%%%%%
\section{\textsc{IMRPhenomXPHM} Analyses with \textsc{bilby}}\label{sec:bilby-imrx}
%%%%%%%%%%%%%%%%%%%%%%%%%%%%%%%%%%

To assess whether differences between \textsc{BayesWave} results and \textsc{bilby} results are due to waveform systematics, we also perform two analyses using \textsc{bilby} and \textsc{IMRPhenomXPHM}: one on subtracted data (Run~\ref{rid:subtracted-bilby-imrx} in Table~\ref{tab:all-analyses-with-settings}), and one using the slow scattering glitch model (Run~\ref{rid:scattering-imrphenomxphm} in Table~\ref{tab:all-analyses-with-settings}).
The analysis on subtracted data found $\chi_{\mathrm{eff}} \leq 0$ at $99.3\%$ credibility, while the analysis marginalizing over the slow scattering model found $\chi_{\mathrm{eff}}\leq 0$ at $99.9\%$ credibility.
From this we conclude that the observed differences between \textsc{bilby} and \textsc{BayesWave} are due to the choice of glitch model, rather than the choice of waveform approximant.

%%%%%%%%%%%%%%%%%%%%%%%%%%%%%%%%%%
\section{Alternate Slow Scattering Glitch Priors}\label{sec:appendix-effect-of-glitch-priors}
%%%%%%%%%%%%%%%%%%%%%%%%%%%%%%%%%%

To test our assumption that it is appropriate to use the slow scattering model with five scattering arches, we also perform a test using four scattering arches with uniform amplitude and targeted modulation frequency priors (Run~\ref{rid:scattering-uniform-targeted-n4} in Table~\ref{tab:all-analyses-with-settings}).
This result finds $\chi_{\mathrm{eff}}\leq 0$ with $\sim 100\%$ credibility, indicating that the inclusion of an arch around $42\, \si{Hz}$ does not alter the conclusions of this work.

The slow scattering model under the physically expected range of modulation frequencies $f_{\rm mod}\sim \mathcal{U}(0.05 - 0.3) \, \si{Hz} $ results in arches that are too extended in time to match the $36\,$Hz excess power morphology.
We explore what values of $f_{\rm mod}$ are required in order to impact $\chieff$ inference, with an analysis that employs a uniform amplitude prior and a maximum modulation frequency of $5 \, \si{Hz}$ (Run~\ref{rid:scattering-uniform-unphysical} in Table~\ref{tab:all-analyses-with-settings}). 
We recover a tri-modal structure favoring $f_{\mathrm{mod}}= 1.5 \, \si{Hz}$ and a less negative $\chieff$, with $\chieff<0$ at 77.1\% credibility.
However, $f_{\mathrm{mod}}= 1.5 \, \si{Hz}$ is $10$ times larger than the scattering surface motion witnessed by the channel \texttt{L1:SUS-ETMX\_L2\_WIT\_L\_DQ}.
Such a result would presume the existence of some alternative source of frequency modulated phase noise, either due to another scattering surface driven at a different frequency, or some non-scattering mechanism, coincidentally aligned in time and frequency with the known scatterer.
While we cannot rule out the existence of such a source, there is no physical motivation to presuppose its existence.
We instead use this analysis to emphasize the conclusion from the \textsc{BayesWave} study, namely that sufficiently flexible glitch models allow for a wider range of possibilities. 

%%%%%%%%%%%%%%%%%%%%%%%%%%%%%%%%%%
\section{Frequency Bin $\chi^2$ Test}\label{sec:chisq-methodology}
%%%%%%%%%%%%%%%%%%%%%%%%%%%%%%%%%%

\begin{figure}
    \centering
    \includegraphics[width=\columnwidth]{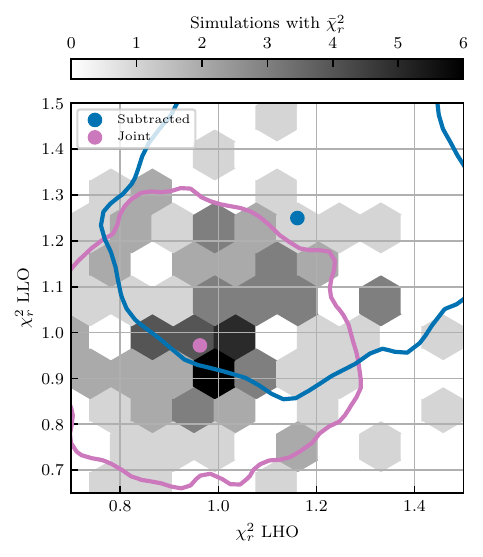}
    \caption{Distribution of $\chi^2_r$ in both detectors for the glitch-subtracted analysis (blue; Run~\ref{rid:subtracted-fd-fL20-fH20}) and the glitch-marginalized analysis (purple; Run~\ref{rid:scattering-uniform-targeted}). The colormap corresponds to the distribution of $\bar{\chi}^2_r$ from simulated signals consistent with GW191109, Runs~\ref{rid:simulated-full-band-start}-\ref{rid:simulated-full-band-end}.
    Dots denote the distribution mean and contours denote the 90\% level.
    For the reference distribution, we histogram the $\bar{\chi}^2_r$ values in LHO and LLO from each simulation.
    }
    \label{fig:chi-square-analysis-and-simulations}
\end{figure}

Tests which assess the Gaussianity of data~\cite{Macas:2023zdu,  Yamamura:2024vka} may be applied to residual data after glitch and signal subtraction, but these do not address whether the signal model is capturing any glitch power. 
In this Appendix we instead consider the frequency bin $\chi^2$ test as employed by search algorithms~\cite{Allen:2004gu, Usman:2015kfa, Davis:2020nyf}.
Qualitatively, it assesses tension between the signal waveform and the data over the entire frequency band, and hence measures deviations due both to model misspecification and to distribution of power not characteristic of a CBC, e.g., a glitch.  

For each posterior waveform, we divide the frequency band into $p$ bins of equal optimal SNR.
If the data are consistent with the sum of the waveform in question and Gaussian noise, then the matched-filter SNR will also be evenly distributed over these bins.
For the $j$th bin, the matched-filter SNR $\rho_{\mathrm{mf}, j}$ will deviate from the mean $\rho_{\mathrm{mf}}$
\begin{equation}
    \Delta \rho_{\mathrm{mf}, j} = \rho_{\mathrm{mf}, j} - \frac{\rho_{\mathrm{mf}}}{p}\,.
\end{equation}
The statistic
\begin{equation}
    \chi^2 = p \sum_{j=1}^{p} |\Delta \rho_{\mathrm{mf}, j}|^2\,, 
\end{equation}
is distributed according to a $\chi^2$ distribution with $2p - 2$ degrees of freedom under Gaussian noise\cite{Allen:2004gu}.\footnote{Two degrees of freedom correspond to the real and imaginary components in each bin, while two are removed since deviations must sum to zero in each of the real and imaginary components.}
The normalized statistic
\begin{equation}
    \chi^2_r = \frac{\chi^2}{2p - 2}\,,
    \label{eq:chir}
\end{equation}
will then have an expected value of 1.
Deviations indicate that the data might not be solely described by the waveform plus Gaussian noise, likely due to a glitch.
We compute $\chi^2_r$ for each GW191109 signal posterior sample on data where the corresponding glitch posterior sample has been subtracted.
We denote the mean statistic over posterior samples as $\bar{\chi}^2_r$.
We then compare against corresponding results from the simulated signals of Sec.~\ref{sec:injection-results}. 
The reason we compare against simulations rather than directly the frequentist expectation for Eq.~\eqref{eq:chir} is that the distribution over the posterior samples is not equivalent to a distribution over many Gaussian noise realizations.

In Fig.~\ref{fig:chi-square-analysis-and-simulations} we plot the statistic distribution over posterior samples in both detectors for the glitch-subtracted analysis of Run~\ref{rid:subtracted-fd-fL20-fH20} and the glitch-marginalized analysis with the slow-scattering model of Run~\ref{rid:scattering-uniform-targeted}.
The colormap corresponds to results from simulated signals where we bin the mean statistic $\bar{\chi}^2_r$ for each simulated signal.
Glitch-marginalization results in a statistic whose mean is more closely in accordance with the frequentist expectation value in both detectors ($\bar{\chi}_r^2 = 0.96$ and  $\bar{\chi}_r^2 = 0.97$ in LHO and LLO respectively) than glitch-subtraction ($\bar{\chi}_r^2 = 1.16$ and $\bar{\chi}_r^2=1.24$ in LHO and LLO). 
Compared to the simulated signals,
glitch-marginalization results in $\bar{\chi}^2_r$ more extreme than that of 41\% (45\%) of simulations in LHO (LLO), while the glitch-subtracted result has a $\bar{\chi}^2_r$ more extreme than 81\% (90\%) of simulations in LHO (LLO).
To produce a meta-statistic, we use Fisher's method~\cite{fisher1928statistical} to compute the likelihood of these statistics occurring together, assuming that the p-values are uncorrelated.
This creates another $\chi^2$ statistic, this time with two degrees of freedom per detector.
For the glitch-marginalized result, we obtain 2.25, corresponding to a $p$-value of 0.69, while for the glitch-subtracted results we have 7.93, giving a $p$-value of 0.09. 
Consistent with expectations, glitch-marginalization results in residuals that are more consistent with Gaussian noise after removing the glitch and signal reconstruction. 

\bibliography{references}% Produces the bibliography via BibTeX.

\end{document}